%% file: science_all.tex
\documentclass[12pt]{article}
\usepackage{upgreek,amsmath,amssymb,graphicx,color}
\usepackage{scicite}
\usepackage{time}

\newcommand{\e}[1]{\times 10^{#1}}
\newcommand{\msun}{M$_\odot$}
\def\ergs {erg s$^{-1}$}
\def\ni {$^{56}$Ni}
\def\kms {km~s$^{-1}$}
\usepackage[normalem]{ulem}
\usepackage{url}

\input{science_macros.tex}

\newenvironment{sciabstract}{%
\begin{quote} \bf}
{\end{quote}}

\newcounter{lastnote}

\title{A Type Ia supernova at the heart of superluminous transient SN 2006gy}

\author
{Anders Jerkstrand$^{1,2*}$, Keiichi Maeda$^{3,4}$, and Koji S. Kawabata$^5$\\
\\
\normalsize{$^1$ Max-Planck-Institute for Astrophysics,}\\
\normalsize{Karl-Schwarzschild-Stra\ss e 1, 85748 Garching, Germany.}\\
\\
\normalsize{$^2$ The Oskar Klein Centre, Department of Astronomy, Stockholm University,}\\
\normalsize{Albanova 10691, Stockholm, Sweden.}\\
\\
\normalsize{$^3$ Department of Astronomy, Kyoto University,}\\
\normalsize{Kitashirakawa-Oiwake-cho, Sakyo-ku, Kyoto 606-8502, Japan.}\\
\\
\normalsize{$^4$ Kavli Institute for the Physics and Mathematics of the Universe, The
University of Tokyo,}\\
\normalsize{5-1-5 Kashiwanoha, Kashiwa, Chiba 277-8583, Japan.}\\
\\
\normalsize{$^5$ Hiroshima Astrophysical Science Center, Hiroshima University,}\\
\normalsize{1-3-1 Kagamiyama, Higashi-Hiroshima, Hiroshima 739-8526, Japan.}\\
\\
\normalsize{*Corresponding author. Email: anders.jerkstrand@astro.su.se}
}

\date{}

\begin{document}

\baselineskip24pt

\maketitle

\begin{sciabstract}
Superluminous supernovae radiate up to 100 times more energy than normal supernovae. 
The origin of this energy and the nature of their stellar progenitors are poorly understood.
We identify neutral iron lines in the spectrum of one such transient, SN 2006gy, and show that they require a large mass of iron ($\gtrsim$0.3 \msun) expanding at 1500 \kms. We demonstrate that a model of a standard Type Ia supernova hitting a shell of circumstellar material produces a light curve and late-time iron-dominated spectrum that match SN 2006gy. In such a scenario, common envelope evolution of the progenitor system can synchronize envelope ejection and supernova explosion and may explain these bright transients.
\end{sciabstract}

\noindent Superluminous supernovae (SNe) are a rare type of astrophysical explosion that emit large amounts of energy, more than can be explained by standard powering mechanisms. One of the first observed such supernovae was SN 2006gy, which showed narrow hydrogen lines (Type IIn) indicating interaction with a circumstellar medium (CSM). It radiated about $10^{51}$ erg in a few months \cite{Smith2007,Ofek2007}. Proposed mechanisms to produce such a transient include large amounts of radioactivity in a pair-instability supernova (PISN) \cite{Smith2007}, collision between a core-collapse supernova (CCSN) and a Luminous Blue Variable-like eruption \cite{Smith2007b}, and a pulsational pair-instability explosion \cite{Woosley2007}. However, the nature of SN 2006gy remains unclear and disputed. 

\begin{figure}
\includegraphics[width=1\linewidth]{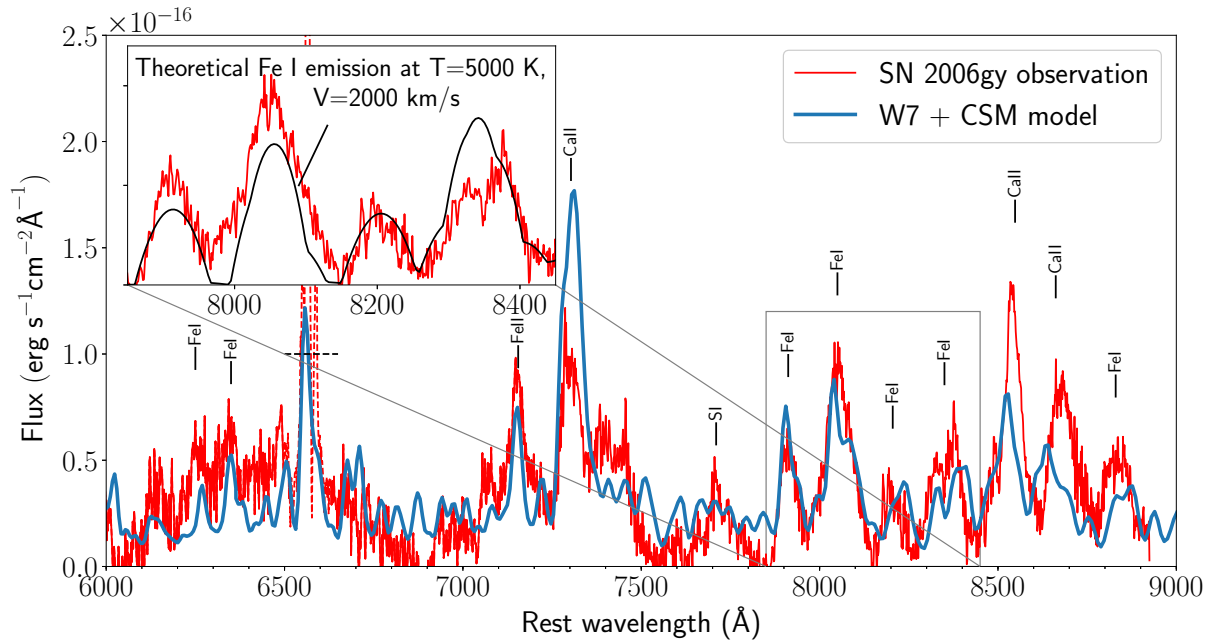}
\caption{\textbf{Observed spectrum of SN 2006gy at +394d (red) compared to standard Ia model (W7) with hydrodynamic velocities reduced by a factor of 7 (blue)}. The model also has 3 $M_\odot$ of CSM mixed with the Ia ejecta. The black dashed line shows the upper limit on H$\alpha$ emission from the supernova. The inset shows a zoom-in on the lines we identify as Fe~\textsc{\small I} at 7900-8500 \AA. The black line shows a theoretical model of emission from 0.5 \msun~of Fe at 5000 K, scaled to the same distance as SN 2006gy. }
\label{fig:compW7}
\end{figure}

A spectrum of the supernova at +394 days post-explosion \cite{Kawabata2009} revealed a set of emission lines around 8000 \AA~that could not be identified. Figure \ref{fig:compW7} shows this spectrum, after removal of echoes (light from earlier epochs reflected by circumstellar dust) \cite{SM}. By searching atomic line lists, we determined that these lines all coincide with low-excitation, strong transitions in Fe~\textsc{\small I} \cite{SM}. 

These lines are predicted by emission line models for slow-expanding supernova ejecta \cite{Jerkstrand2018}. They arise from  the z$^7$D multiplet of Fe~\textsc{\small I} at 2.4 eV above the ground state, which is excited by thermal electron collisions at typical supernova temperatures of a few thousand kelvin. Most supernovae have, however, too little neutral iron and too high expansion velocities to produce these lines in their spectrum. In addition to these Fe~\textsc{\small I} lines, the spectrum of SN 2006gy shows lines from Ca~\textsc{\small II} and Fe~\textsc{\small II} and is thus dominated by heavy elements likely produced in explosive oxygen and silicon fusion. The FWHM (Full-Width-at-Half-Maximum) of these iron and calcium lines are $\sim 1500$ km s$^{-1}$, which corresponds to the characteristic expansion velocity of the gas at +394d.

\begin{figure}
\centering
\includegraphics[width=0.8\linewidth]{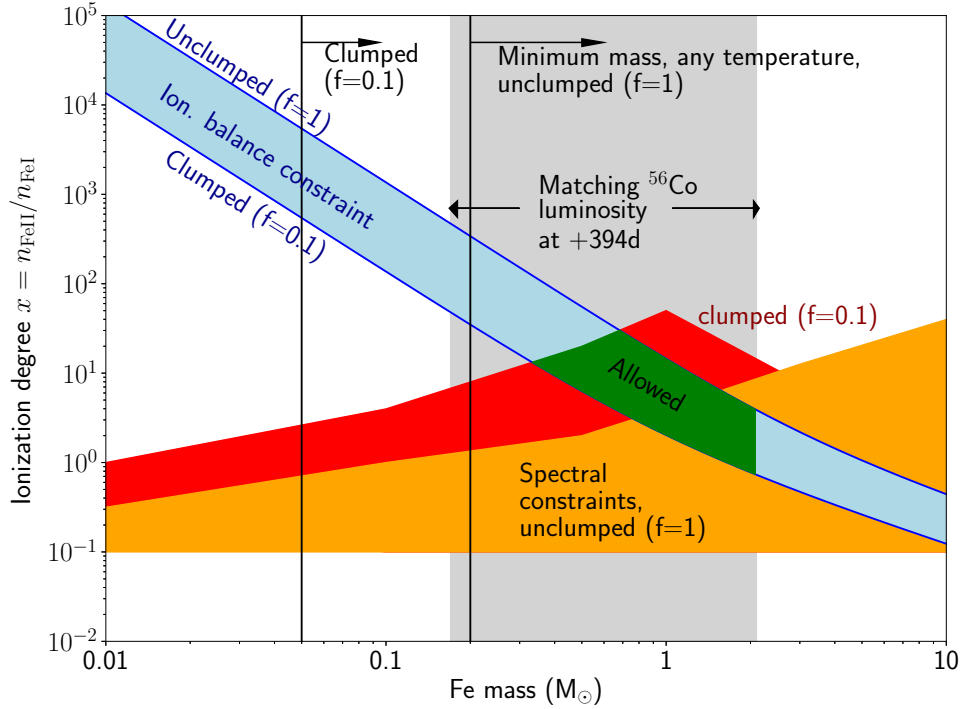}
\caption{\textbf{Allowed domains for iron mass and ionization degree}. The vertical lines show the minimum Fe masses needed to reproduce the Fe~\textsc{\small I} lines at +394d for any temperature and ionization. The blue domain outlines where ionization balance holds. The orange/red domains outline where the observed Fe~\textsc{\small II}/ Fe~\textsc{\small I} and Fe~\textsc{\small I} cluster emission ratios are reproduced, for unclumped (orange) and clumped (red) cases. The gray regime outlines where the luminosity is reproduced by the remaining amount of $^{56}$Co, if the iron comes from $^{56}$Ni/$^{56}$Co decay. The regime where all constraints are fulfilled at $f \gtrsim 0.1$ is marked green; it constrains 0.3 \msun~$\lesssim$  $M(\rm Fe) \lesssim$~2.1 \msun.}
\label{fig:B}
\end{figure}

To obtain constraints on the iron producing this emission, we calculated a grid of iron (Fe~\textsc{\small I} and Fe~\textsc{\small II}) emission models with the spectral synthesis code \textsc{\small SUMO} \cite{Jerkstrand2011} varying iron mass, temperature, ionization, and clumping (degree of compression compared to a uniform distribution) \cite{SM}. Small masses of iron ($\lesssim 0.1$ $\rm M_\odot$ (solar masses)) cannot produce the observed luminosity for any physical conditions (Fig. \ref{fig:B}) . To both fulfil ionization balance and reproduce the observed emission ratio between Fe~\textsc{\small I} and Fe~\textsc{\small II} lines, another constraint $M(\mbox{Fe}) \gtrsim$ 0.3 $\rm M_\odot$ can be derived, assuming a filling factor (inverse of clumping) between 0.1-1 \cite{SM}. Lower masses give a too high ionization state and results in emission mainly from Fe~\textsc{\small II} and Fe~\textsc{\small III} rather than Fe~\textsc{\small I}. The iron mass limit holds also under exploration of smaller filling factors \cite{SM}. A large mass of iron is therefore inferred, likely arising from decayed $^{56}$Ni (through the intermediary $^{56}$Co), the main product of explosive silicon fusion. 

At +394d after explosion, SN 2006gy was about 100 times fainter compared to previous observations at +200d. A fundamental property of a localized CSM is that the shock will traverse the CSM on a time-scale $230~\mbox{days} \left(R/10^{16}~\mbox{cm}\right)\textbf/\left(v_\text{shock}/5000~\mbox{km s}^{-1}\right)$, where $R$ is the radius and $v_{\rm shock}$ is the shock speed. Similar drops in brightness have been seen in other luminous Type IIn supernovae \cite{Fransson2014,Inserra2016}.  In its second and third year after explosion, SN 2006gy became dominated by an echo with slower decay than either interaction or radioactive powering \cite{Fox2015}.

The amount of initial radioactive $^{56}$Ni needed to match the estimated luminosity of the supernova at +394d is 0.5 $\rm M_\odot$ \cite{SM}. Figure 1 (in-set) shows the theoretical emission spectrum of 0.5 $\rm M_\odot$ of Fe~\textsc{\small I} at 5000 K, scaled to the same distance as SN 2006gy, which reproduces the observed Fe~\textsc{\small I} lines. These strong iron lines in SN 2006gy are difficult to reconcile with several previously suggested models in which there is no \ni~production, for example the collisions of pulsational pair instability shells \cite{Woosley2007}.

Core-collapse supernovae (arising when the core of a massive star collapses to a neutron star or black hole) produce much less \ni, typically $\lesssim$0.1 
\msun\cite{Muller2017,Prentice2019}, although a small fraction, virtually all in the broad-lined Ic class, have inferred \ni~ masses $\gtrsim$0.3 \msun~\cite{Mazzali2001}. Such an engine for SN 2006gy can however be excluded on two grounds. First, for a CCSN to produce 0.5 \msun\ of \ni, the explosion energy has to be over $10^{52}$ erg \cite{Heger2010}. Because wind-driven mass loss and pair instability pulsations limit the final mass of the supernova progenitor to about ten solar masses, these supernovae expand fast (6,000-12,000) \kms~as confirmed by late-time spectroscopic observations \cite{Mazzali2001}. For such a supernova to reach a velocity of 1500  \kms~after a few hundred days, the ejecta must have been strongly decelerated by a massive CSM, with associated re-radiation of the bulk of the original kinetic energy ($\sim 10^{52}$ erg). The observed radiated energy in SN 2006gy is an order of magnitude lower at $10^{51}$ erg \cite{SM}, preventing any self-consistent CCSN scenario. Second, CCSN ejecta are dominated by oxygen, with strong [O~\textsc{\small I}] lines after a few hundred days, of which SN 2006gy shows none.

Two model scenarios can explain a $^{56}$Ni mass of $\sim$0.5 \msun~expanding with 1500 km s$^{-1}$ at 400d - a pair-instability explosion of a $\sim$90 \msun~He core \cite{Heger2002}, and a Type Ia supernova (the thermonuclear explosion of a white dwarf (WD)) decelerated by strong circumstellar interaction. The ejecta mass needed to trap the radioactive decay gamma-rays (that transfer the decay energy to heat) at 400d is 1.8 \msun~(setting the optical depth $\tau_\gamma = \kappa_\gamma \rho R = 1$, where $\kappa_\gamma$ is the gamma-ray opacity and $\rho$ is the density), and the gamma-rays therefore mainly power the supernova ejecta rather than the CSM in both cases.

We calculated model spectra for both scenarios with \textsc{\small SUMO}, and found good agreement for both, as they have similar core structures. Figure \ref{fig:compW7} shows spectra using the W7 explosion model \cite{Nomoto1984, Iwamoto1999} with all velocities in the hydrodynamic model reduced by a factor of 7 to mimic the slowdown due to CSM interaction (leading to higher densities at any given time). We mixed the ejecta with a few solar masses of CSM material, however the spectrum was not sensitive to this \cite{SM}. This W7+CSM model reproduces the Fe~\textsc{\small I} lines, the [Ca~\textsc{\small II}] doublet, and the only ionized iron line seen, [Fe~\textsc{\small II}] 7155 \AA. The Ca~\textsc{\small II} triplet at 8500-8700 \AA~is underproduced, possibly because the Ca-rich region 
is not compact enough; higher density favours a stronger triplet line. 

\begin{figure}
\includegraphics[width=0.48\linewidth]{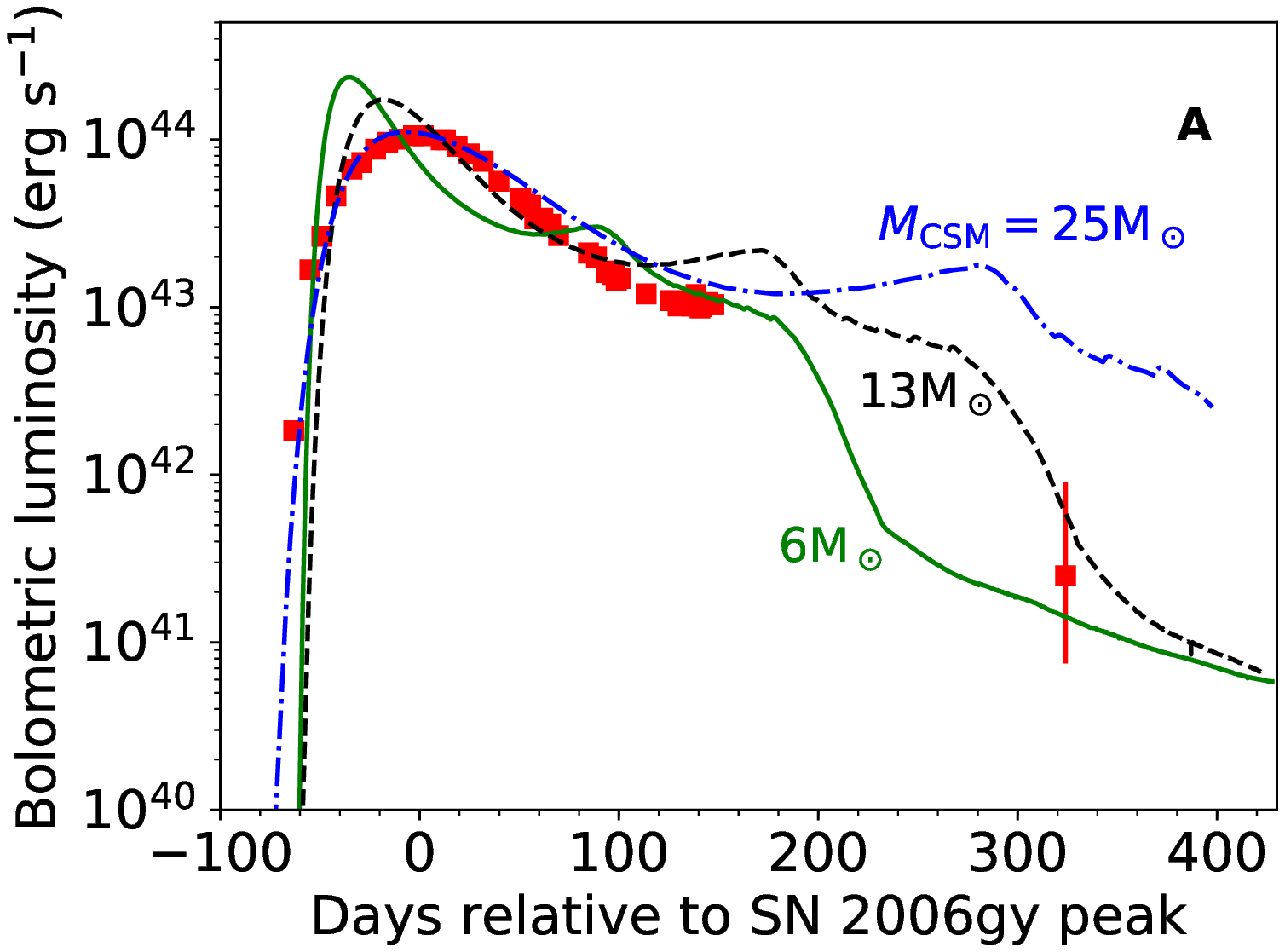} 
\includegraphics[width=0.50\linewidth]{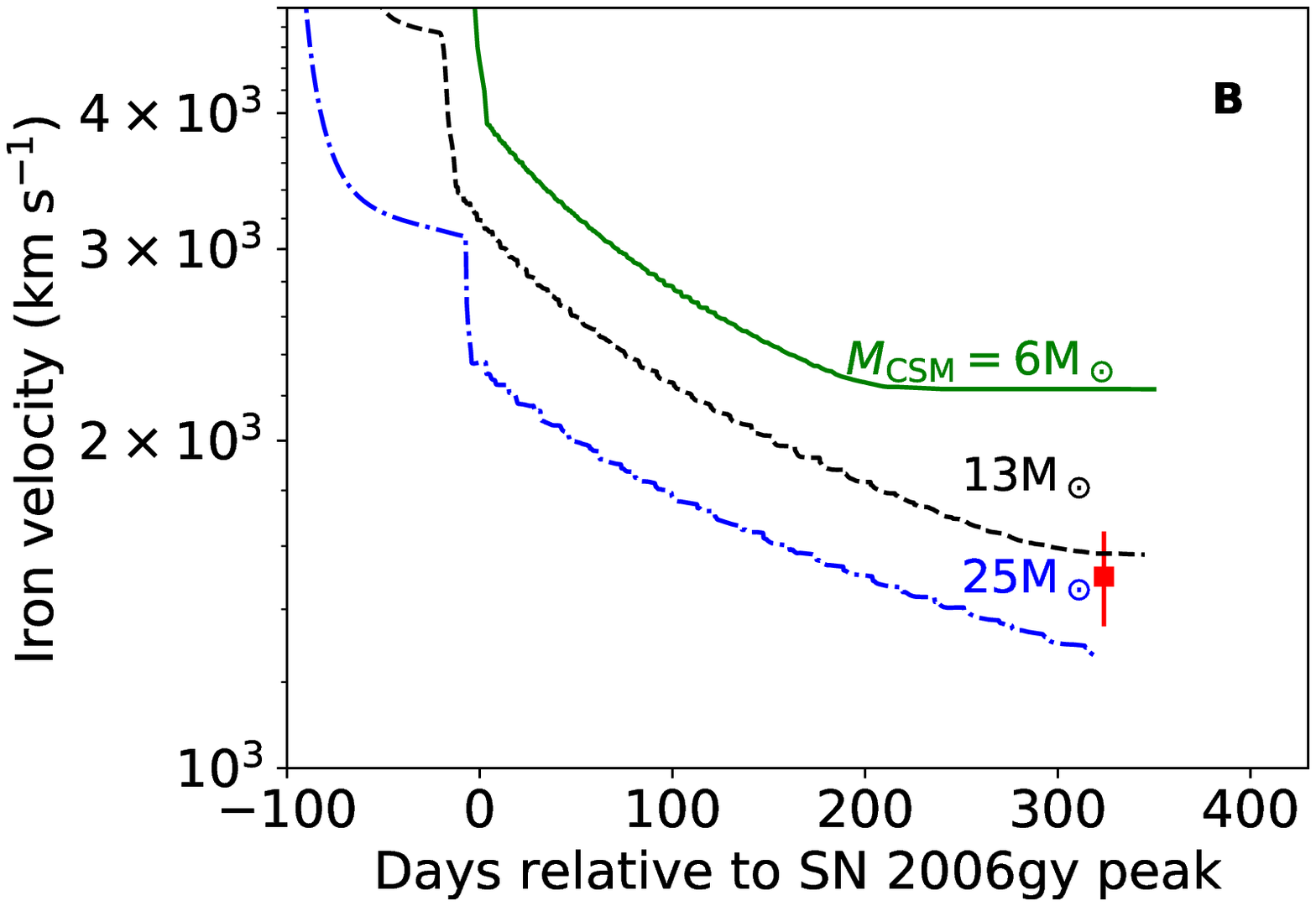}
\caption{\textbf{Type Ia-CSM light curve models compared to SN 2006gy (red points), showing bolometric luminosity (emission integrated over all frequencies, panel A) and velocity of the Ia ejecta (at $M=0.5$~\msun, panel B)}. This iron only becomes observable in the nebular phase spectrum at +394d. All models have outer CSM radius $R_{\rm CSM}=8\e{15}$ cm. 
The green curve is representative of models with too small CSM mass ($\lesssim$ 6 \msun); the light curve peaks too narrow and bright, and the deceleration is insufficient. The blue curve is representative of models with too much CSM ($\gtrsim 25$ \msun) ; interaction powers the light curve for too long and the slowdown is too severe. The black model represents a case of CSM mass (13 \msun) where all properties are reproduced.}
\label{fig:C}
\end{figure}

The nebular-phase degeneracy between Type Ia and PISN models can be broken by considering the earlier phases of the supernova. We calculated the total amount of light emitted by SN 2006gy using all the spectral and photometric data available in the literature \cite{Smith2007,Agnoletto2009,Smith2010}. We obtain $9\times 10^{50}$ erg, close to that expected in the strong interaction limit of a Type Ia supernova where a large fraction of the kinetic energy of $(1-2)\times 10^{51}$ erg is converted to radiation \cite{SM}. Some previous estimates of this number were a factor 2-3 higher, but were based either on single-band data with an assumed bolometric correction \cite{Smith2007}, or extrapolated blackbodies with high ultraviolet(UV)/blue flux \cite{Smith2010}. Such UV/blue emission is often blocked by line opacity in supernovae and the spectra of SN 2006gy show such behaviour \cite{SM}. We used the radiation hydrodynamic code \textsc{\small SNEC} \cite{Morozova2015} to calculate light curves arising when a standard Ia SN ejecta (W7), as well as PISN ejecta, collide with a dense H-rich CSM. The resulting light curves for the Ia case match SN 2006gy if a CSM mass of about 10 \msun~is present (Fig. \ref{fig:C}). Pair-instability supernovae, on the other hand, produce light curves in strong disagreement with observations (Fig. \ref{fig:gridbig}).

Inspection of the Ia-CSM hydrodynamic models shows that the ejecta are decelerated to 1500 km s$^{-1}$ following interaction with a CSM with properties suitable for reproducing the light curve. This matches the observed velocities of the Fe~\textsc{\small I} lines at +394d. The Type Ia explosion energy, $1.3\e{51}$ erg for the standard scenario \cite{Iwamoto1999}, is accounted for with about $3\e{50}$ erg still in kinetic energy at 400d ($\sim$15 $\rm M_\odot$ at 1500 km s$^{-1}$, both ejecta and CSM expand with this asymptotic velocity), and the rest radiated. The ``Type Ia-CSM'' hypothesis thus matches all observables. 
This scenario has been previously proposed for SN 2006gy \cite{Ofek2007}, but was then largely forgotten as most analyses focused on massive star progenitors.

From the CSM extension and velocity, the CSM material must have been ejected between 10-200y before the supernova explosion. A candidate scenario to explain this is common envelope evolution of a binary progenitor system, in which a white dwarf spirals into a giant or supergiant companion star. This could causally link the processes of envelope ejection and a merger with the core of the other star, producing the explosion. 
Such synchronization by common envelope evolution has previously been discussed in other contexts \cite{Chevalier2012}.
The inspiral process has been shown to robustly transfer energy and angular momentum from the orbit to the common envelope, and eject most or all of this, while the orbital separation shrinks by a factor 100 or more \cite{Terman1994,Taam2000}. 

The ejection time-scale in SN 2006gy matches the time-scales for common envelope ejection obtained in simulations; $\sim$10y for red giants engulfing WDs \cite{Terman1994}, and 2-200y for more massive red supergiants (RSGs) \cite{Taam2000}. 
The released orbital energy for a WD of mass $M_{\rm WD}$ spiralling in towards a companion with core mass $M_{\rm core}$ and radius $R_{\rm core}$ is 
\begin{equation}
4\times 10^{48} \left(\frac{M_{\rm core}}{\rm M_\odot}\right) \left(\frac{M_{\rm WD}}{\rm M_\odot}\right) \left(\frac{R}{\rm R_\odot}\right)^{-1} \mbox{erg},
\end{equation} 
where $\rm R_\odot$ is the solar radius.
This is sufficient to unbind 10 $\rm M_\odot$ of envelope material in
an extended star (binding energy $4\times 10^{48}$ erg for an average $R=100~\rm R_\odot$) and also account for the kinetic energy of the ejected envelope ($10^{48}$ erg for 100 km s$^{-1}$). 
It is less clear how the two cores merge and explode. These steps are rarely explored in inspiral simulations due to the computational difficulties, although some results have shown that less evolved giants merge more easily \cite{Taam2000}. 
Material may also form a disc around the two cores that could drive the last merging steps \cite{Kashi2011}. 

A similar scenario may explain Type IIa supernovae, a rare class which have spectra of Type Ia at early times but later transition to Type IIn (but much less luminous than SN 2006gy). One suggestion laid forth is the common envelope ejection in a merger of a WD and an Asymptotic Giant Branch (AGB) star \cite{Livio2003}. Such a scenario was criticized on the grounds that the final merger would have to occur by gravitational waves, which would take much longer than decades or centuries \cite{ChugaiY2004}. However, the last stages of common envelope evolution are not well understood, so that conclusion may be premature. 

It is possible that SN 2006gy is an extreme example of the Ia-CSM family, with higher CSM mass located closer to the supernova compared to other cases. This would be 
more efficient at converting kinetic energy to radiation, over a shorter time scale, leading to the extreme luminosity. It also led to strong ejecta deceleration that trapped gamma-rays and produced the distinct narrow Fe lines after a few hundred days. Other IIa supernovae show longer-lasting interactions with a more extended CSM, which would not slow the expanding core sufficiently to produce a distinct signature from the inner ejecta at late times. 

Other superluminous Type IIn SNe such as SN 2006tf \cite{Smith2008b}, SN 2008fz \cite{Drake2010} and SN 2008am \cite{Chatzo2011} share several similarities with SN 2006gy. The total radiated energy in these events is also around $10^{51}$ erg, so some may also represent a Type Ia SN exploding in a massive common envelope-ejected CSM. These other objects were however much further away, and a similar signature as the +394d spectrum of SN 2006gy was not observable for them, with attempts at late-time observations yielding either no detections or still ongoing interaction through broad hydrogen lines \cite{Smith2008b,Drake2010,Chatzo2011}.

\bibliographystyle{Science}
\bibliography{scibib}

\noindent \textbf{Acknowledgements:} We thank N. Ivanova, R. Pakmor, R. Iaconi and J. Grumer for discussion. 
We thank M. Rampp and L. Stanisic at the Max Planck Computing and Data Facility for computing assistance. 
We also thank the referees and the editor for much useful feedback that improved the manuscript.

\noindent \textbf{Funding:} A.J. acknowledges funding by the European Union's Framework Programme for Research and Innovation Horizon 2020 under Marie Sklodowska-Curie grant No. 702538 and European Research Council (ERC) Starting Grant No. 803189, as well as support from the Swedish Research Council and Swedish National Space Board. K.M. acknowledges support from JSPS Kakenhi grants 18H05223, 18H04585, and 17H02864. K. K. acknowledges support from JSPS grants 18H03720 and 17H06363.

\noindent \textbf{Author contributions:} A. J. identified the Fe~\textsc{\small I} lines in SN 2006gy and performed the spectral modelling. K. M. performed the radiation hydrodynamic modelling. K. K. and A. J. performed the data calibration. All authors contributed to the analysis and paper writing.

\noindent \textbf{Competing interests:} The authors declare no competing interests.

\noindent \textbf{Data materials and availability:} The spectrum shown in Fig. 1 is available on the WiseRep database (\url{https://wiserep.weizmann.ac.il/system/files/uploaded/general/2006gy\_2007-09-18\_Subaru\_FOCAS\_None\_0.xy}). The \textsc{\small SNEC} code is available at https://stellarcollapse.org/SNEC. The SUMO code was developed by A. Jerkstrand, C. Kozma and C. Fransson. All input and output files for the \textsc{\small SUMO} and \textsc{\small SNEC} modelling, and SUMO executables, are available at the Max-Planck-Society data repository EDMOND in collection ``SN2006gy'' \cite{SN2006gy}(\url{https://dx.doi.org/10.17617/3.30}).\\
\\
\noindent{\textbf{SUPPLEMENTARY MATERIALS}}\\
Materials and Methods\\
Supplementary text\\
Figs. S1 to S11\\
Tables S1 to S2\\
References (32-66)

\newpage % ==============================================================
\setcounter{page}{1}
%\section*{
\begin{center}
\huge Supplementary Materials for\\
%\\
\vspace{1cm}
\large A Type Ia supernova at the heart of superluminous transient SN 2006gy\\
\vspace{1cm}
\normalsize Anders Jerkstrand, Keiichi Maeda, Koji Kawabata\\
\vspace{1cm}
\normalsize Correspondance to: anders.jerkstrand@astro.su.se
\end{center}

\renewcommand\thefigure{S\arabic{figure}}    
\setcounter{figure}{0}

\renewcommand\thetable{S\arabic{table}}    
\setcounter{table}{0}

\begin{flushleft}
\textbf{This PDF file includes}:\\
~~~Materials and Methods\\
~~~Supplementary text\\
~~~Figs. S1-S11\\
~~~Tables S1-S2\\
~~~References 31-65\\
\end{flushleft}

\newpage
\subsection*{Materials and Methods}

\subsubsection*{\underline{Data calibration and uncertainty estimates.}}
The extinction towards SN 2006gy is high, with both Na~\textsc{\small I D} lines and spectral energy distribution (SED) comparisons to other IIn SNe giving color excess (difference between observed and intrinsic color) values in the range $E(B-V)=0.5-0.75$ mag\cite{Smith2007,Ofek2007,Agnoletto2009}. We use here $E(B-V)=0.63 \pm 0.15$ mag, the average of the various estimates, which keeps consistency with the value used in \cite{Kawabata2009}. 

Of the three possible background subtractions in \cite{Kawabata2009} (their figure 5), the one with less subtraction is the only one consistent with the $V$-band photometry (21.0 mag of spectrum vs $20.7 \pm 0.4$ mag for photometry). We avoid using the $R$-band for comparison because there is disagreement between this value in \cite{Kawabata2009} ($19.4\pm 0.4$ mag) and the non-detection limits of 19.5 and 20.3 mag in \cite{Smith2008} and \cite{Agnoletto2009}.  We use the $V$-band consistent spectrum, and note that for the Fe~\textsc{\small I} emission lines around 8000 \AA~central to our analysis, there are no major differences between the three extractions (within 10\%). We assess a contribution of $\pm$0.3 mag to the uncertainty due to calibration (based on figure 5 in \cite{Kawabata2009}). We apply a 20\% correction for slit losses (i.e. multiply the spectrum by 1.2), a typical value (the seeing for the night was 0.6$''$).

\begin{figure}
\includegraphics[width=1\linewidth]{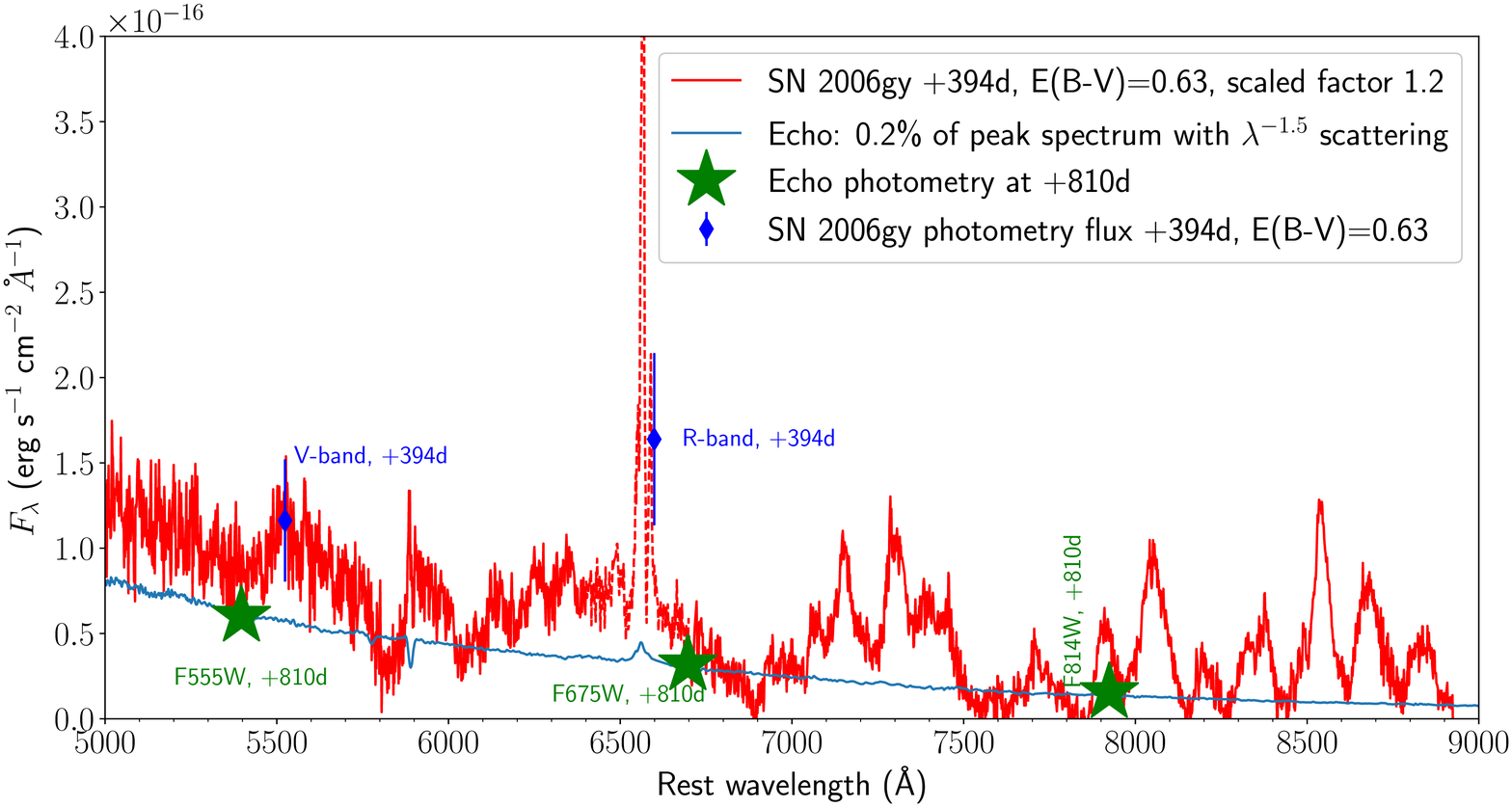}
\caption{\textbf{Spectrum and photometry of the SN 2006gy region at late times (394 and 810d post explosion).} The +394d spectrum from \cite{Kawabata2009} is shown in red, the contemporary photometry \cite{Kawabata2009} as blue diamonds, and the photometry of the echo at +810d in three bands (F555W, F675W, F814W) from \cite{Miller2010} as green stars. The SN peak light spectrum \cite{Smith2010} filtered by a $\lambda^{-1.5}$ scattering law is shown in blue.}
\label{fig:fulldata}
\end{figure}

The SN region produced an echo that was clearly detected in observations at later epochs, well described by the peak light SED filtered by a power law scattering law \cite{Miller2010}. As standard models for echoes predict constant or slowly declining flux levels \cite{Chevalier1986}, the +394d spectrum is likely affected by the same, or very similar, echo. The observed spectrum shows the blue steepening with time which is the hallmark of echoes, and the +810d photometry corresponds to some 2/3 of the flux at +394d, with the SN emission lines still visible with a $\sim$1/3 contribution. We experimented with different power laws and found $\lambda^{-1.5}$ to give best reproduction to the \cite{Miller2010} photometry, a value similar to theoretical estimates \cite{Patat2005}. Fig. \ref{fig:fulldata} shows the late-time spectra and photometry, and the echo model we subtract to obtain the pure SN spectrum. A small amount of continuum flux ($1.5\e{-17}$ erg s$^{-1}$ cm$^{-2}$ \AA$^{-1}$) was added back after the echo subtraction to avoid negative flux levels.

The association of the H$\alpha$ line with the supernova is uncertain as the residual lies as a dim bridge in the 2D spectrum between galactic H$\alpha$ and [N II] 6548, 6583 \AA~\cite{Kawabata2009}. As in \cite{Kawabata2009} we plot H$\alpha$ with a dashed line to indicate this uncertainty whether the line is from the SN or not. In \cite{Smith2010} it is argued that from higher resolution data this H$\alpha$ line may not belong to the SN.

If we take estimates for the uncertainty in distance modulus ($\pm0.15$ mag), extinction ($\pm 0.6$ mag), photometry ($\pm 0.4$ mag) and background spectral subtraction ($\pm 0.3$ mag) in quadrature, the total 1$\sigma$ uncertainty in flux level is $\pm 0.7$ mag, or a flux scaling factor $0.5-1.9$. At 2$\sigma$ (95\% confidence) the scaling factor range is $0.3-3.6$. We use this 95\%\ range in the analysis to determine allowed model fits, i.e. any model that needs to be scaled by a factor within $0.3-3.6$ is deemed viable, whereas models needing more rescaling than this are discarded.

\subsubsection*{\underline{Line identifications and single-zone modelling.}}
Table \ref{table:lines} lists the identified observed lines, with measured half-width zero-intensity (HWZI) values. The lines are broader ($\sim$1500 km s$^{-1}$) than the spectrograph resolution (450 km s$^{-1}$). Thus, these are intrinsic SN velocities only slightly broadened by the spectrograph (an estimate of the supernova expansion velocity is $v_{\rm SN} \approx \sqrt{1500^2 - 450^2} = 1430$ km s$^{-1}$). The identifications are lines from Fe~\textsc{\small I}, Fe~\textsc{\small II} and Ca~\textsc{\small II}. The Fe~\textsc{\small II} and Ca~\textsc{\small II} lines are commonly seen in SN spectra, whereas Fe~\textsc{\small I} lines are not. There is no sign of lighter element lines such as [O I]~\textsc{\small 6300, 6364} \AA~or [C I]~\textsc{\small 8727} \AA. The heavy element identification, and low expansion velocities, therefore suggest an association with the innermost silicon and oxygen burning layers of the supernova.

\begin{table}
\caption{\textbf{Distinct observed lines in the +394d spectrum, with measured central wavelengths in the supernova rest frame, widths (Half-width Zero Intensity HWZI), and identifications (line data from \cite{Kurucz1995})}. The transition level numbers (in energy-order) are listed in parentheses. The S~\textsc{\small I} 7725 identification is speculative.}
\centering
\begin{tabular}{ccl}
\hline
Observed central wavelength  & HWZI  & Identification\\
(\AA)   & (km s$^{-1}$)   & \\
\hline
7150 & 1600 & [Fe~\textsc{\small II}] 7155 (17-6) \\  
7300 & 1800 & [Ca~\textsc{\small II}] 7291,7323 (2-1)\\ 
7710 & $<$2000 & S~\textsc{\small I} 7725 (5-4) \\ 
7905 & 1700 & Fe~\textsc{\small I}] 7912 (21-6)  \\ 
8047 & 2000 & Fe~\textsc{\small I}] 8047 (18-6) + Fe~\textsc{\small I}] 8075 (23-7)  \\ 
8200 & 1400 & Fe~\textsc{\small I}] 8204 (25-8) \\ 
8315 & 1300  & Fe~\textsc{\small I}] 8311 (23-8) + Fe~\textsc{\small I}] 8307 (26-9) \\ 
8360 & 1700 & Fe~\textsc{\small I}] 8349 (18-7) + Fe~\textsc{\small I}] 8382 (25-9)  \\ 
8540 & 1300 & Ca~\textsc{\small II} 8542 (4-2) \\  
8675 & 2000 & Ca~\textsc{\small II} 8662 (3-2)   \\ 
\hline
\end{tabular}
\label{table:lines}
\end{table}

The distinct group of Fe~\textsc{\small I} lines (7912-8349 \AA) come from multiplet z$^7$D, which makes up levels 18, 21, 23, 25 and 26 when energy-ordered. This multiplet has an excitation energy of 2.4 eV, sufficiently low for thermal collisional excitation. The transitions are down to the a$^5$F multiplet producing the cluster of lines around 8000 \AA. These fulfil selection rules for electric dipole transitions, but involve a change in the total spin in LS coupling, and are thus semi-forbidden intercombination lines, with Einstein A-values around $10^2$ s$^{-1}$.

To confirm the Fe~\textsc{\small I} identification, it is important to consider what other lines from Fe~\textsc{\small I} should be expected. From inspection of lines arising from upper levels below 4 eV (excitation beyond this would require unrealistically high temperatures $\gtrsim 10^4$ K), one can see that emission clusters at $\sim$4400 \AA, $\sim$5100 \AA, $\sim$6400 \AA, and $\sim$8200~\AA~ (the identified cluster) are expected. The first two clusters are too blueward to probe in SN 2006gy (the 5100 \AA~one is a borderline case), but the one at 6400 \AA~falls in the observed range at +394d. The observed spectrum does have a broad bump centred on 6400 \AA, with similar luminosity as the 8200 \AA~cluster.

Using the \textsc{\small SUMO} spectral synthesis code \cite{Jerkstrand2011,Jerkstrand2012} a single-zone Non-Local-Thermodynamic-Equilibrium (NLTE) grid for Fe~\textsc{\small I} + Fe~\textsc{\small II} was constructed with the following four parameters allowed to vary:
\begin{itemize}
\item $M(\rm Fe) = 0.01,~0.03,~0.05,~0.07,~0.1,~0.2,~0.3,~0.4,~0.5,~1,~3,~10~\rm M_\odot$
\item Number density ratio $n_{\rm Fe\textsc{\tiny II}}/n_{\rm Fe\textsc{\tiny I}}=0.1,~1,~10,~100$
\item $T = 2000,~3000,~4000,~5000,~6000,~7000,~8000~\rm K$ 
\item Filling factor $f = 0.1,~1$ (fraction of volume filled with material, i.e. inverse of clumping)
\end{itemize}
The grid has in addition constant parameters $t = 400$d, $v_{\rm SN}=2000$ km s$^{-1}$, and $n_{\rm Fe\textsc{\tiny II}} = n_{\rm e}$. 
If CSM is mixed together with the iron, the electron density may deviate from $n_{\rm Fe\textsc{\tiny II}}$ by a factor few but by experimentation we found this not to have any large impact on the results.
The grid calculates NLTE emissivities, with Sobolev escape probabilities, with the aim to outline the physical regimes needed to match the data. More detailed multizone models including line-to-line radiative transfer are considered in the next section.

\noindent We seek the subset of models in this grid that fulfil four key properties of the data:
\begin{enumerate}
\item Luminosity between 7900-8500 \AA~(dominated by the Fe~\textsc{\small I} lines) that matches the observed one within the derived uncertainty scaling of $0.3-3.6$.
\item Luminosity over the whole observed wavelength range (5000-8900 \AA) less than 3.6 times the observed one (no lower limit as it is unknown how much other elements than Fe~\textsc{\small I} + Fe~\textsc{\small II} may contribute).
\item An acceptable ratio between the two Fe~\textsc{\small I} clusters at 6400 and 8200 \AA~(see below).
\item An acceptable ratio between the Fe~\textsc{\small I} and Fe~\textsc{\small II} emission~(see below).
\end{enumerate}
Figure \ref{fig:grid3} shows that, for the unclumped scenario, an iron mass $\gtrsim 0.2$ \msun~is needed to produce the observed Fe~\textsc{\small I} lines, for any temperature and density combination (i.e. just constraints (1) and (2)). If one plots the same figures for $f=0.1$ (factor 10 clumping), this is lowered to $M(\mbox{Fe}) \gtrsim 0.05$ \msun. Much more clumping than factor 10 is unlikely for radioactive material which tends to stay expanded due to its internal heating \cite{Herant1991,Basko1994}. One-dimensional hydrodynamic models can give artificially high compression factors, a well known limitation \cite{Chen2016}, and are therefore not suited to estimate this.

\begin{figure}
\centering
\includegraphics[width=0.49\linewidth]{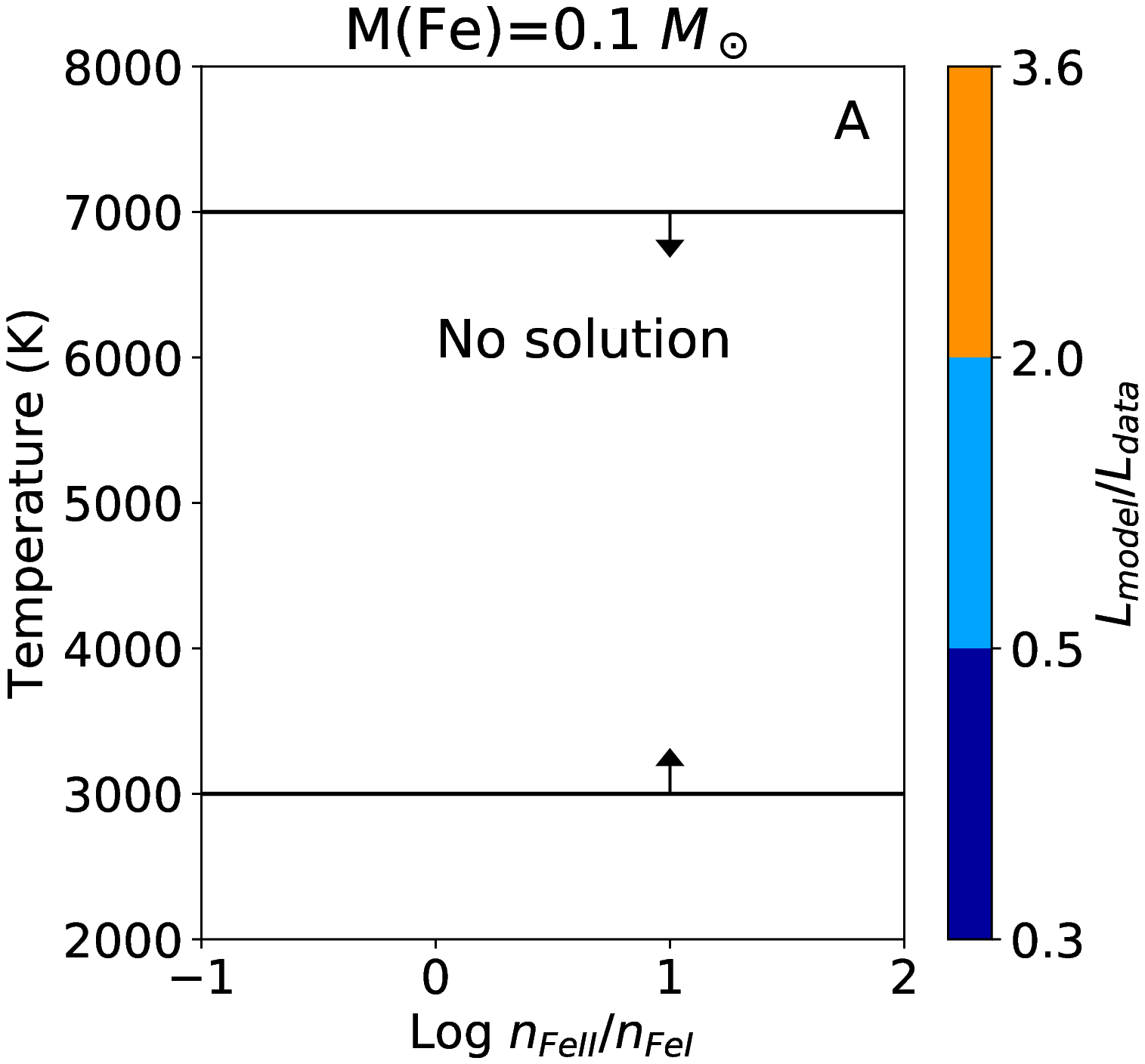}
\includegraphics[width=0.49\linewidth]{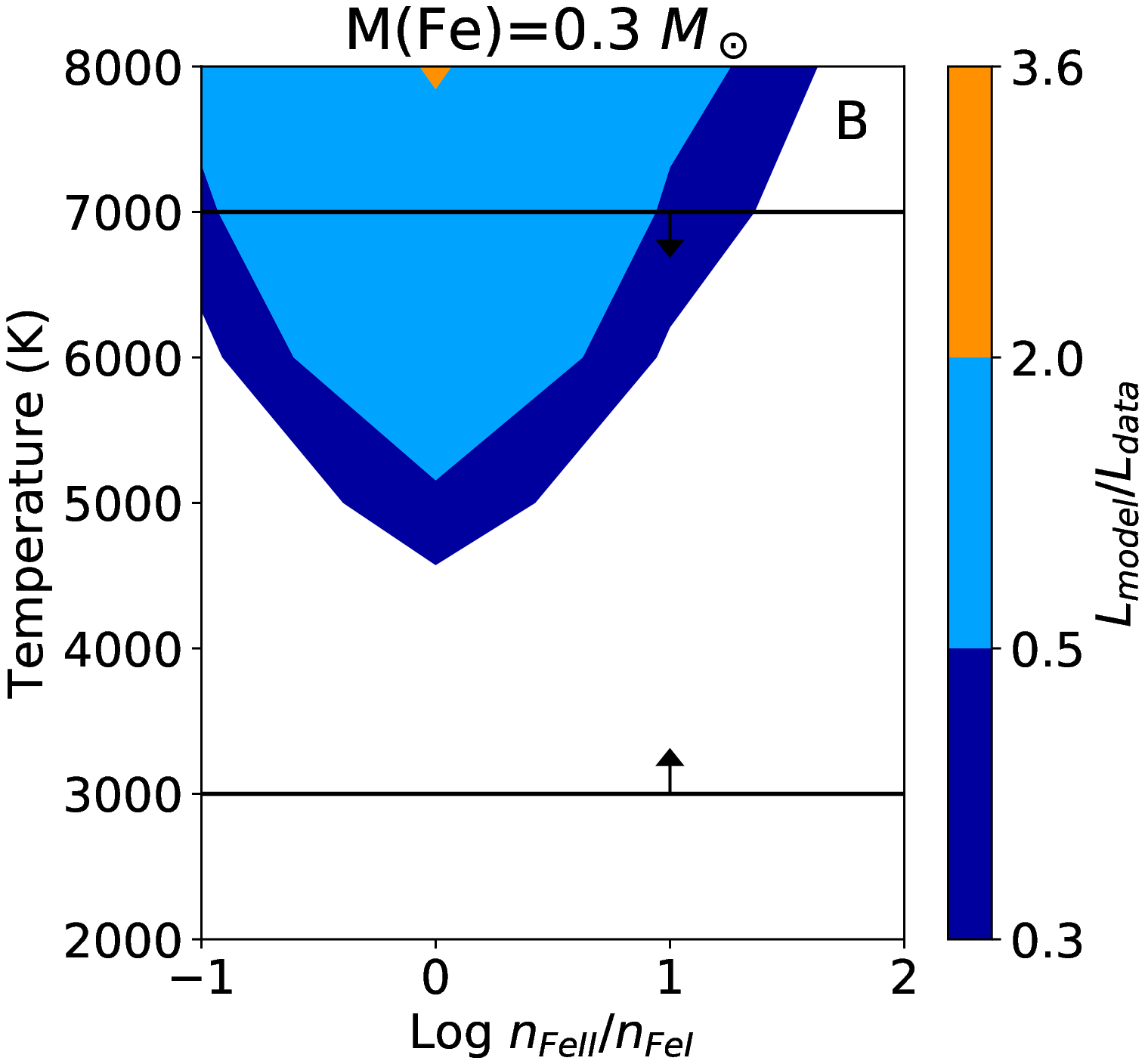}\\
\includegraphics[width=0.49\linewidth]{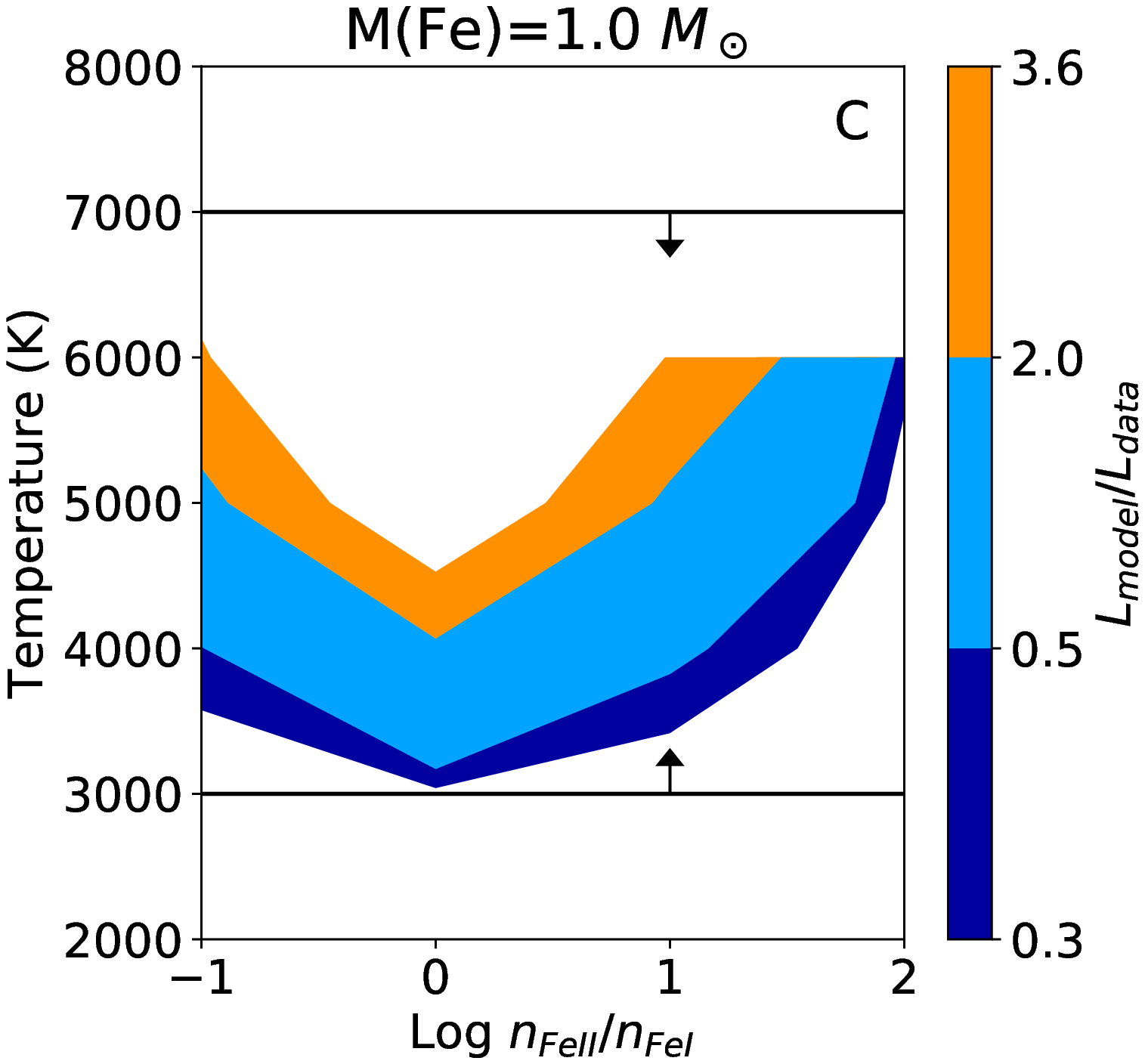}
\includegraphics[width=0.49\linewidth]{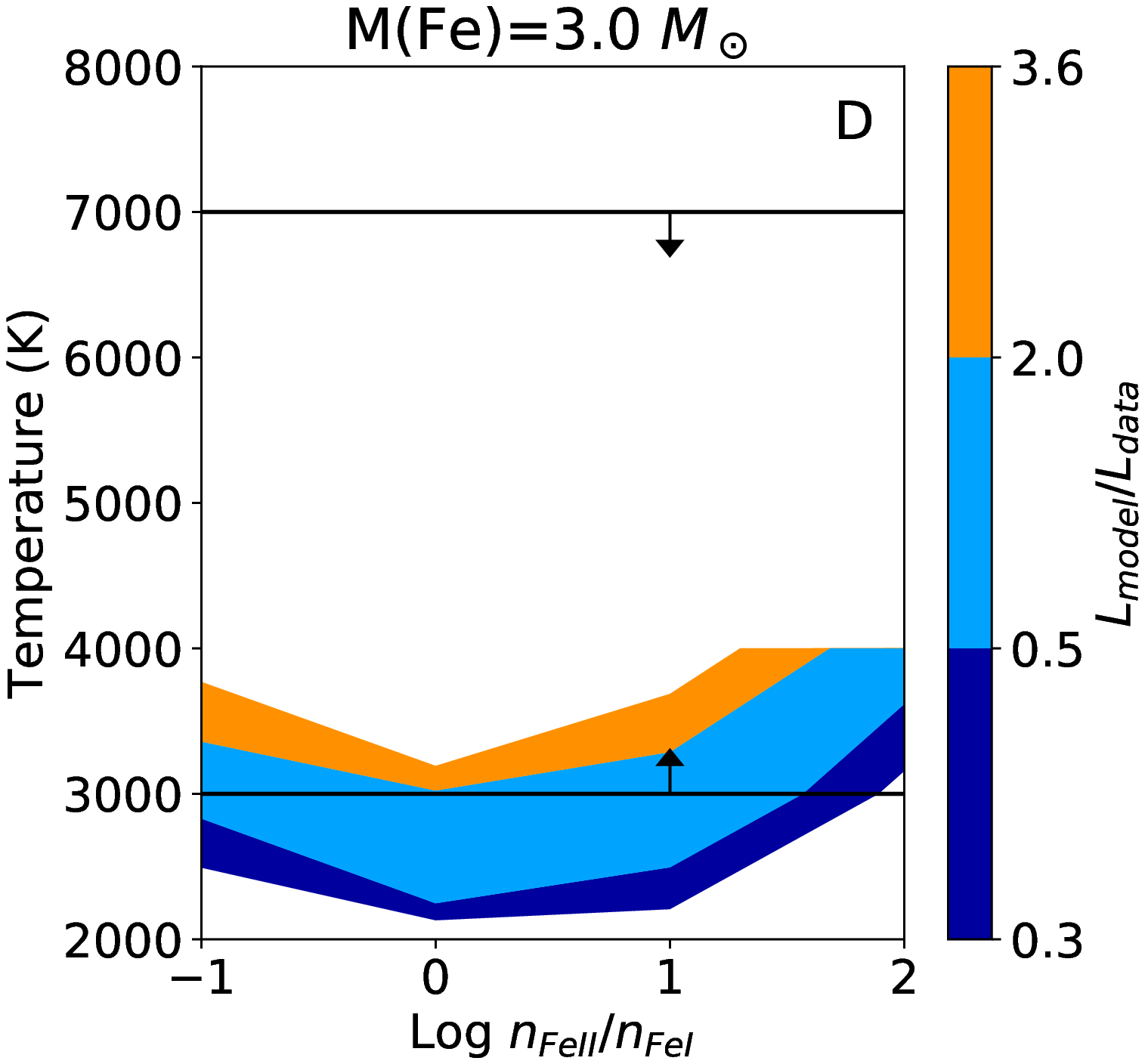}
\caption{\textbf{The ratio of NLTE model luminosity (7900-8500 \AA) relative to observed one, for single-zone models with $f=1$}. The subpanels show Fe masses of 0.1 \msun~(A), 0.3 \msun~(B), 1.0 \msun~(C) and 3.0 \msun~(D). The specified iron mass is uniformly distributed over a spherical region with $v_{\rm SN}=2000$ km s$^{-1}$, and this (together with $t=400$d) sets also the total number density. The temperature (y-axis) and ionization state of the iron (x-axis) are then varied, under the constraint $n_{\rm e} = n_{\rm FeII}$. Regions where the 5000-8900 \AA~flux exceeds 3.6 times the observed are blanked out even if the 7900-8500 is within the observed range. The temperature range indicated by line ratios (3000-7000 K) is marked.}
\label{fig:grid3}
\end{figure}

\begin{figure*}
\centering
\includegraphics[width=1\linewidth]{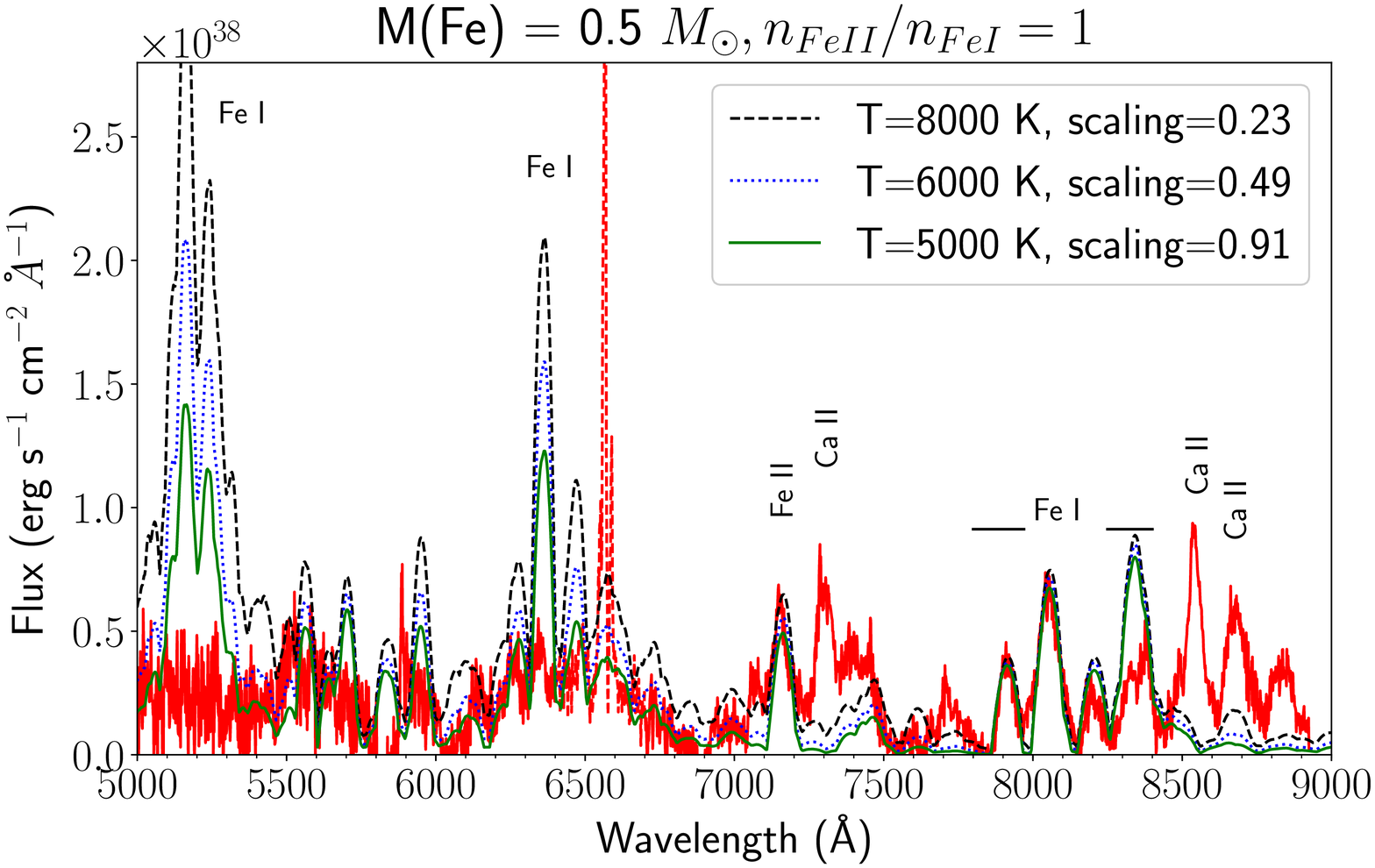}
\caption{\textbf{Single-zone iron NLTE model spectra ($M(\mbox{Fe})=0.5$ $\rm M_\odot$ and $f=1$) at different temperatures compared to SN 2006gy (red, echo subtracted)}. The models have been rescaled to reproduce the Fe~\textsc{\small I} cluster at 7900-8400 \AA~(scaling factors written in legend). Models with too high temperature give overproduction of the blue Fe~\textsc{\small I} clusters ($\sim 5300$ \AA\ and $
\sim 6400$ \AA).}
\label{fig:plotter3}
\end{figure*}

\noindent \textbf{Temperature constraints from ratio of Fe~\textsc{\small I} clusters}. Figure \ref{fig:plotter3} shows three spectra at $M(\mbox{Fe}) = 0.5$ \msun, with different temperatures, varying the scaling to fit the Fe~\textsc{\small I} cluster at 7900-8500 \AA. As expected from the discussion above, there are clusters also at $\sim$5200 \AA~and $\sim$6400 \AA~from Fe~\textsc{\small I}.
Models with high temperatures, $T \gtrsim 6000$ K, have problems with overproduction of these blue clusters, and inspection of the grid shows that this is an issue independent of the iron mass and ionization state. It means that many regimes with small iron masses $M < 0.5$ \msun~are problematic from Fe~\textsc{\small I} line ratios as they require $T \gtrsim 6000$ K and an associated overproduction in the blue that is too large for even the quite large uncertainty in the data as well as possible radiative transfer effects to explain (photons at shorter wavelengths have in general higher chance of being absorbed). To get some estimate of the degree of line blocking we compared the W7 model with and without line opacity switched on. When rescaled to have the same luminosity in the 8200 \AA~cluster, the model with lines off was a factor $\sim$1.5-2 brighter in the 6300 \AA~cluster and a factor $\sim$5 brighter in the 5200 \AA~cluster. This indicates that line blocking is not sufficient to allow $T > 7000$ K. 
We quantitatively impose this constraint by putting a roof at 7000 K for allowed solutions outlining the ``spectral constraints'' regions in Fig. \ref{fig:B}. In the same way, temperatures below $\sim$3000 K give too little 6400 \AA~emission, as it arises from a higher multiplet than the 8400 \AA~lines, and this defines a floor. This temperature constraint (3000-7000 K) is used in the next section.

\noindent \textbf{Ionization constraint from Fe~\textsc{\small II} to Fe~\textsc{\small I} ratio.} A constraint on $n_{\rm FeII}/n_{\rm FeI}$ can be put from the observed ratio of Fe~\textsc{\small II} and Fe~\textsc{\small I} lines. There is a clear observed [Fe~\textsc{\small II}] 7155 line with luminosity equalling 50\% of Fe~\textsc{\small I}] 8047, the strongest Fe~\textsc{\small I} line. Fig. \ref{fig:B} shows the $n_{\rm FeII}/n_{\rm FeI}$ regimes where the model ratio between [Fe~\textsc{\small II}] 7155 and Fe~\textsc{\small I}] 8047 lies within 0.5-2 times the observed value (0.5), for any temperature in the range 3000-7000 K. In these regions, line ratios between Fe~\textsc{\small I} clusters and between Fe~\textsc{\small I} and Fe~\textsc{\small II} are therefore both fulfilled. No luminosity constraints are imposed here, just line ratios.

\noindent \textbf{Ionization balance.} Consider now the ionization balance. The ionization rate per volume is
\begin{equation}
\Gamma n_{\rm FeI} = \frac{L}{N_{\rm FeI} \chi_{\rm FeI}} \psi_{\rm ion,FeI} n_{\rm FeI} = \frac{L}{V_0 f \chi_{\rm FeI}} \psi_{\rm ion,FeI}^{\rm 0}(1-x_e) \tag{S1}
\end{equation}
where $\Gamma$ is the ionization rate per particle, $L$ is the bolometric luminosity (our estimate is $2.5\e{41}$ erg s$^{-1}$ at +394d), $N_{\rm FeI}$ is the total number of Fe~\textsc{\small I} atoms, $\chi_{\rm FeI}$ is the ionization potential of Fe~\textsc{\small I}, $V_0$ is the total volume (for $f=1$), $n$ are number densities, and $x_e$ is the electron fraction. The $\psi_{\rm ion,FeI}$ coefficient is the fraction of deposited energy going into ionization of Fe~\textsc{\small I}, which scales roughly with the fraction of atoms in the Fe~\textsc{\small I} state, which is represented by the $1-x_e$ factor. 

\noindent The recombination rate per volume is $n_e^2 \alpha(T) = n^2 x_e^2 \alpha(T)$, where $\alpha(T) \sim 1.5\times 10^{-12}$ cm$^3$ s$^{-1}$ for Fe~\textsc{\small I} at a few thousand K \cite{Nahar1997}. Then, ionization balance gives
\begin{equation}
\frac{L}{V_0 f \chi_{\rm FeI}} \psi_{\rm ion,FeI}^{\rm 0}(1-x_e) = \frac{M_{\rm Fe}^2 \times \left(2\times 10^{33}~\mbox{g}\right)^2}{V_0^2 f^2 \bar{A}^2 m_p^2} \alpha(T) x_e^2 \tag{S2}
\label{eq:ionbal}
\end{equation}
where $\bar{A}=56$ is the atomic weight of iron. Defining the dimensionless quantity
\begin{equation}
A = \frac{L \psi_{\rm ion,FeI}^{\rm 0} \bar{A}^2 m_p^2 V_0}{\chi_{\rm FeI} \alpha(T)\times \left(2\times 10^{33}~\mbox{g}\right)^2} \tag{S3}
\end{equation}
which has value $A=10.6$ for our standard values ($L=2.5\e{41}$ erg s$^{-1}$, ~$\chi_{\rm FeI}= 7.9$ eV, $V_0=4\pi/3\times (2000~\mbox{km s}^{-1}\times 400\mbox{d})^3$, and $\psi_{\rm ion,FeI}=1/3$ (a typical fraction of deposited energy going to ionization when Fe~\textsc{\small I} is abundant, \cite{Kozma1992}).
Eq. \ref{eq:ionbal} has solution
\begin{equation}
x_e = -\frac{Af}{2M_{\rm Fe}^2} + \sqrt{ \frac{A^2 f^2}{4 M_{\rm Fe}^4} + \frac{Af}{M_{\rm Fe}^2}} \tag{S4}
\end{equation}
Plotting $x=n_{\rm e}/n_{\rm FeI} = x_{\rm e}/(1-x_{\rm e})$ (Fig. \ref{fig:B}) shows that low iron masses ($\lesssim 0.1$ \msun) are unavoidably associated with high ionization, $x \gtrsim 10^2$ and little Fe~\textsc{\small I}. This is in conflict with the strong observed Fe~\textsc{\small I} lines and low ionization inferred from the observed Fe~\textsc{\small I} to Fe~\textsc{\small II} emission ratios, requiring $x \lesssim 10$. A low iron mass cannot reproduce either the luminosity, the Fe~\textsc{\small I} emission, or the Fe~\textsc{\small I} to Fe~\textsc{\small II} ratio in SN 2006gy. Should the SN go into freeze-out so ionization balance breaks down (but this usually occurs later \cite{Fransson1993,Fransson2015}), the ionization will be higher, making this constraint even stronger. We have also neglected photoionization, which will push the constraint in the same direction.

Finally, we clarify that the upper limits to $x$ plotted in Fig. 2 are set by the lower limit on the temperature (3000 K) and not the upper limit (7000 K), and the resulting requirement $M(\rm Fe) > 0.3$ \msun~is therefore not sensitive to the 7000 K limit. Neither is the Fe mass very sensitive to the 3000 K limit, because at those low temperatures the minimum masses of Fe creep up to $\gtrsim 0.5$ \msun~from pure luminosity constraints (a constraint not explicitly visible in Fig. 2 as the temperature axis is compressed). This means that 0.3 \msun~is a robust lower limit.\\

\subsubsection*{\underline{Two further indications of at least 0.5 \msun\ of explosive burning ashes.}} % ========================
With a large Fe mass identified spectroscopically, which likely comes from decayed $^{56}$Ni, a natural scenario to investigate is that at least the later phases of SN 2006gy are powered by radioactive decay of this $^{56}$Ni/$^{56}$Co. There are two further, independent results that point to a high Fe mass:
\begin{enumerate}
\item At +394d, the 5000-9000 \AA~output of the SN is $1.0\times10^{41}$ \ergs~(subtracting the echo). Applying a correction for missing flux in the ultraviolet/near-infrared/mid-infrared (UV/NIR/MIR), which is a factor $\sim 2.5$ (see Section ``Energy budget'' below) at the last epoch, the estimated bolometric luminosity is $2.5\e{41}$ erg s$^{-1}$. This corresponds to 0.56 \msun\ of \ni\  produced in the explosion. Thus there is self-consistency between the mass inferred from spectral properties and the late-time luminosity. This would therefore favor a $M\approx 0.5$ \msun~solution out of the otherwise allowed $M \gtrsim 0.3$ \msun~limit inferred from spectra. However, one must keep in mind here the large uncertainty factor of 0.3-3.6 for the +394d spectral flux levels, which means masses between 0.2-2.1 \msun~becomes formally allowed by this argument. While the diffusion peak of SN 2006gy is interaction-powered, several arguments (disappearance of H$\alpha$, lack of any X-ray or radio emission at late times) indicate that this interaction abates after $\sim$200d \cite{Smith2008, Smith2010}, and a switch to radioactivity as the power source is consistent with the data. The qualitative change in spectra from narrow H-lines arising from outer powering from circumstellar interaction to broader iron-group lines from central powering is further circumstantial evidence for this.

\item It is unlikely that this powering is due positrons (carrying 3.5\% of the decay energy), as it would require a total \ni\ mass of 15 \msun\, and such a hefty explosion would give much higher expansion velocities (also with any reasonable circumstellar interaction). If a large fraction of gamma rays (carrying 96.5\% of decay energy) would escape the \ni\ region, they would also provide some powering of overlying layers, which is not observed. Thus, it appears most plausible that the powering comes from gamma rays trapped in the \ni-rich inner region. To trap the gamma rays at an epoch of 400d (gamma-ray optical depth $\tau_\gamma \gtrsim 1$), a zone mass of $\gtrsim$1.8 \msun~is needed in a uniform sphere model with $V=1500$ \kms ($M=4\pi/3 V^2 t^2 /\kappa_\gamma$, where $\kappa_\gamma=0.03$ cm$^2$ g$^{-1}$ is the gamma ray opacity). 
Supernova ejecta can be optically thick to gamma rays while optically thin in the optical; only line opacity matters in the optical and this quickly becomes small after a few months of expansion \cite{J11thesis}.
In a thin shell scenario, the radial optical depth for such a mass drops to 0.3, but as many photons now travel along non-radial directions the average path length is longer, largely offsetting this (we obtain a factor 2.5 longer average path in simulations), and the trapping remains high. As no lines of O, C, He or H are seen (apart from an uncertain H$\alpha$), it is reasonable to assume this mass must be heavy elements (Fe, Si, S, Ca). Thus, a scenario where 1-2 solar masses of \ni-rich gas is powering itself becomes self-consistent; there is both enough power and enough self-trapping, two independent physical constraints.
\end{enumerate}

\subsubsection*{\underline{Multi-zone modelling.}}
The multi-zone modelling uses the standard setup of \textsc{\small SUMO}. We focus here on description of the Ia modelling. For PISN spectral formation see \cite{Dessart2013} and \cite{Jerkstrand2016}.

An uncertainty in the set-up is the extent to which the Ia ejecta is mixed with the CSM shell. We explored different assumptions here. With no mixing, some fraction of the gamma-rays, about 1/3 but depending on the exact morphology, is deposited in the overlying CSM material. Such a component gives emission mainly in quasi-continuum, Ca~\textsc{\small II} triplet lines and H$\alpha$. H$\alpha$ formation is complex, with dependencies on density and radiative transfer effects such as Lyman line overlaps and modelling this in detail is difficult. It is also possible that the outer CSM shell is fragmented and would not trap as much gamma rays energy as in 1D models. In our standard model where the Ia ejecta is mixed together with a few solar masses of CSM (which would be the inner parts of the $\sim$ 10 \msun~total) almost no gamma rays enter outerlying CSM regions. We added also some He (0.5 \msun) to this in-mixed component as such material is present in both AGB and supergiant stars. H$\alpha$ does not emerge particularly strong from the in-mixed CSM, with high Balmer optical depth and iron line blocking damping it. Virtually all emission lines are made by the material of the Ia ejecta, with H and He emitting mainly in the UV, which fluoresces into the optical and NIR to increase the quasi-continuum level. The CSM material does not contribute much to the cooling. 

\subsubsection*{\underline{Light curve modelling.}}
We performed radiation hydrodynamic simulations for a set of the model parameters describing the SN Ia ejecta colliding with dense circumstellar media (CSM), using the radiation hydrodynamic code \textsc{\small SNEC} (Supernova Explosion Code \cite{Morozova2015}). The initial conditions consist of the expanding SN Ia ejecta (the W7 model \cite{Nomoto1984,Iwamoto1999}) and a stationary CSM. For the ejecta, the W7 structure (both in density and composition) is homologously expanded ($R=Vt$) until 1 day after explosion, and this structure is then mapped onto the numerical grid. Outside the SN ejecta, up to a given maximum radius, the CSM is attached. We focus on the CSM density distribution arising from a steady wind $(\rho \propto r^{-2}$); this distribution produces a shape of the light curve compatible with that seen in SN 2006gy (Fig. \ref{fig:C}) for suitable parameters. The goal here is to demonstrate rough reproduction of luminosity, diffusion time scale, and ejecta deceleration, and we expect different configurations (e.g. shells) to give some quantitative but no qualitative differences. We refer the reader to \cite{Moriya2013, Dessart2015, Noebauer2016} for such investigations. One limitation of \textsc{\small SNEC} is the assumption of Local Thermodynamic Equilibrium, although in the scenarios explored here matter is compressed into dense shells so this should be reasonable. Another, probably more important limitation, is that of gray radiative transfer as radiation and matter may be out of equilibrium in interacting supernovae (see e.g. \cite{Dessart2015}). This would likely act to underestimate the opacity and thus overestimate the CSM mass.

Our models form a two-parameter sequence characterizing the nature of CSM; the mass loss rate ($\dot M$) and the maximum CSM radius ($R_{\rm CSM}$). Adopting the wind velocity of the mass loss process ($v_{\rm w}$) as $\sim 100$ km s$^{-1}$ (this can be determined from the observed narrow lines in the spectrum \cite{Smith2007}), the duration of the pre-SN mass ejection episode is then $t_{\rm loss} = R_{\rm CSM} / v_{\rm w} \sim 15 \ {\rm years} \times (R_{\rm CSM}/5 \times 10^{15} \ {\rm cm}) \times \left(v_{\rm w}/100~\rm km~s^{-1}\right)^{-1}$. The total mass of the CSM is $M_{\rm CSM} = \dot M \times t_{\rm loss} \sim 15~\rm M_{\odot}\times(\dot M/1~\rm M_\odot \ {\rm yr}^{-1})\times(R_{\rm CSM}/5 \times 10^{15} \ {\rm cm}) \times \left(v_w/100~\rm km~s^{-1}\right)^{-1}$. In our simulation grid, the ranges of the parameters are (0.25 -- 1.5) $\rm M_\odot$ yr$^{-1}$ for $\dot M$, and (2 -- 17) $\times 10^{15}$ cm for $R_{\rm CSM}$. For the CSM composition, the solar abundance is assumed. 

\begin{figure}
\includegraphics[width=1\linewidth,trim=0mm 0mm 0mm 0mm,clip]{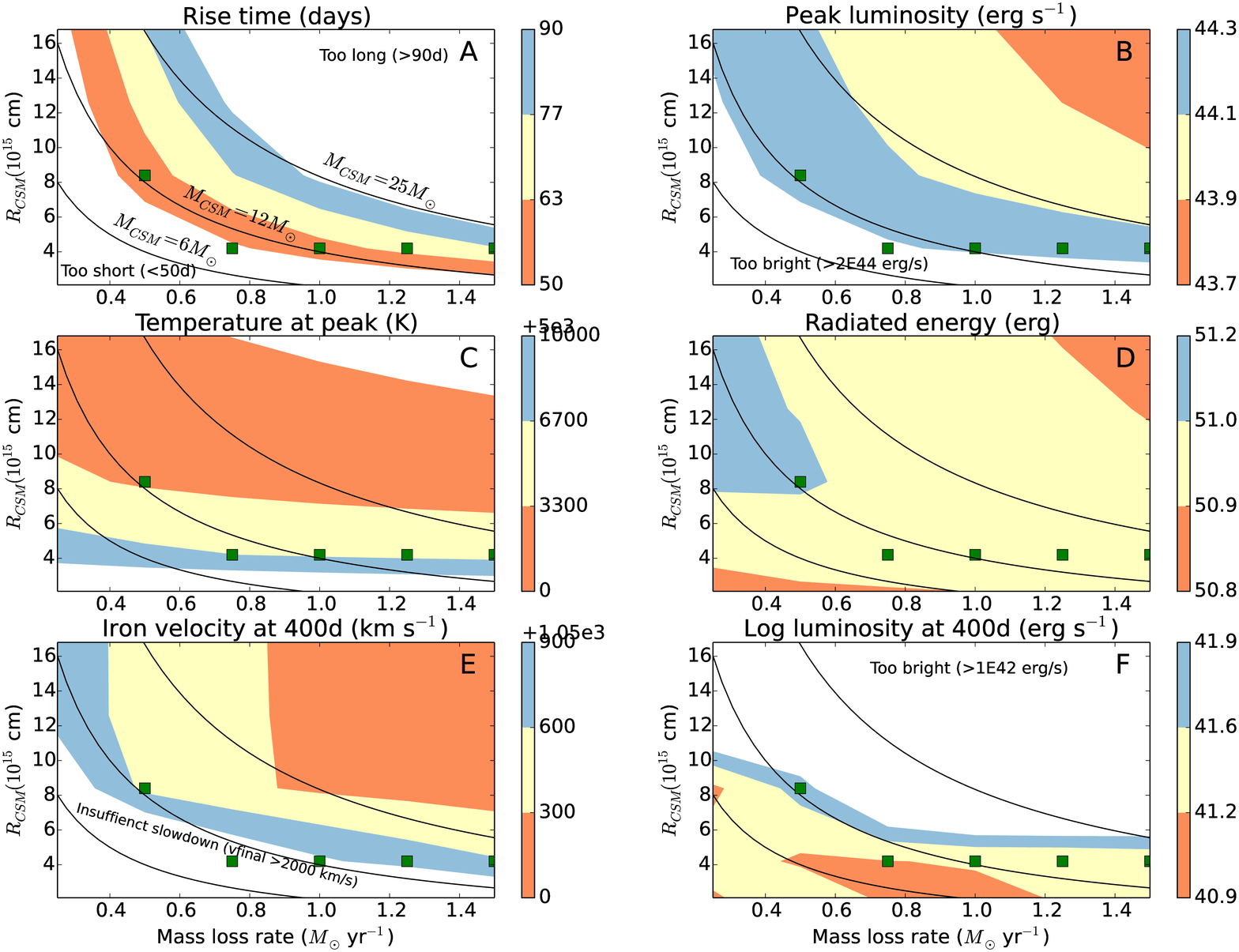}
\caption{\textbf{Properties of the Type Ia-CSM light curve models}. The panels show rise time (A), peak luminosity (B), peak temperature (C), total radiated energy (D), iron velocity (E) and 400d luminosity (F). Allowed domains to be in agreement with SN 2006gy are colored. The specific models fitting all six quantities are marked with green squares. The three black contour lines trace, from left to right, CSM masses of 6, 12 and 25 \msun.}
\label{fig:SNICgrid}
\end{figure}

Typical bolometric light curves computed for a few selected models are shown in Fig. \ref{fig:C}. Here the explosion epoch for each model is chosen to give the overall best fit to the data. The evolution of the velocity at mass coordinate $M = 0.5$ \msun, taken as a representative ejecta velocity, is also shown there. Figure \ref{fig:SNICgrid} shows the rise time, peak bolometric luminosity, peak temperature, radiated energy, iron velocity at 400d, and luminosity at 400d. 

Rise-time is mainly governed by the CSM mass, and  $10-20$ \msun~give the best agreements with the data. Such masses give also the right peak luminosities, whereas lower values are somewhat too bright at peak. The peak temperature and radiatied energies put no strong constraints within the parameter range explored; in all cases is the CSM massive enough that most the of the SN kinetic energy is converted to radiation (see also the simulations in \cite{vanMarle2010} (hydrodynamics with optically thin cooling) and \cite{Dessart2015} (radiation hydrodynamics), where less extreme $M_{\rm CSM}/M_{\rm ejecta}$ ratios of up to 3 were investigated, but even in this regime the conversion efficiency reached 65-70\%. \cite{Moriya2013} apply a free parameter for possible reduction of this conversion factor due to lateral motions, but this is currently not underpinned as multi-D results \cite{vanMarle2010} show this effect to be small. 
See also \cite{Dessart2016,Chugai2016,Sorokina2016} for other simulations in this regime of high $M_{\rm CSM}/M_{\rm ejecta}$ ratio.
The \textsc{\small SNEC} simulations show that the energy extraction efficiency approaches the theoretical limit of 0.94/0.98 for gas pressure/radiation dominated strong shocks \cite{Dyson1997}, respectively, as $M_{\rm CSM}/M_{\rm ejecta} \rightarrow \infty$. For the slow-down and powering at 400d, too low CSM masses do not provide enough deceleration, whereas too high masses still provide too strong interaction at 400d. The models that fulfil all six observational checks are marked with green squares in Fig. \ref{fig:SNICgrid}- they lie close to the $M_{\rm CSM} = 12$ \msun~line. The required extension of the CSM is about $(4-8) \times 10^{15}$ cm, and the mass loss rate is $0.5-1.5~\rm{M_\odot}$ yr$^{-1}$. The ejection time-scale is $t_{\rm loss} \sim 25$y, although allowing for that some (moderate fraction) of the measured CSM velocities may have arisen in the acceleration \cite{Dessart2015}, this sets just a lower limit.

\begin{figure}
\includegraphics[width=0.49\linewidth]{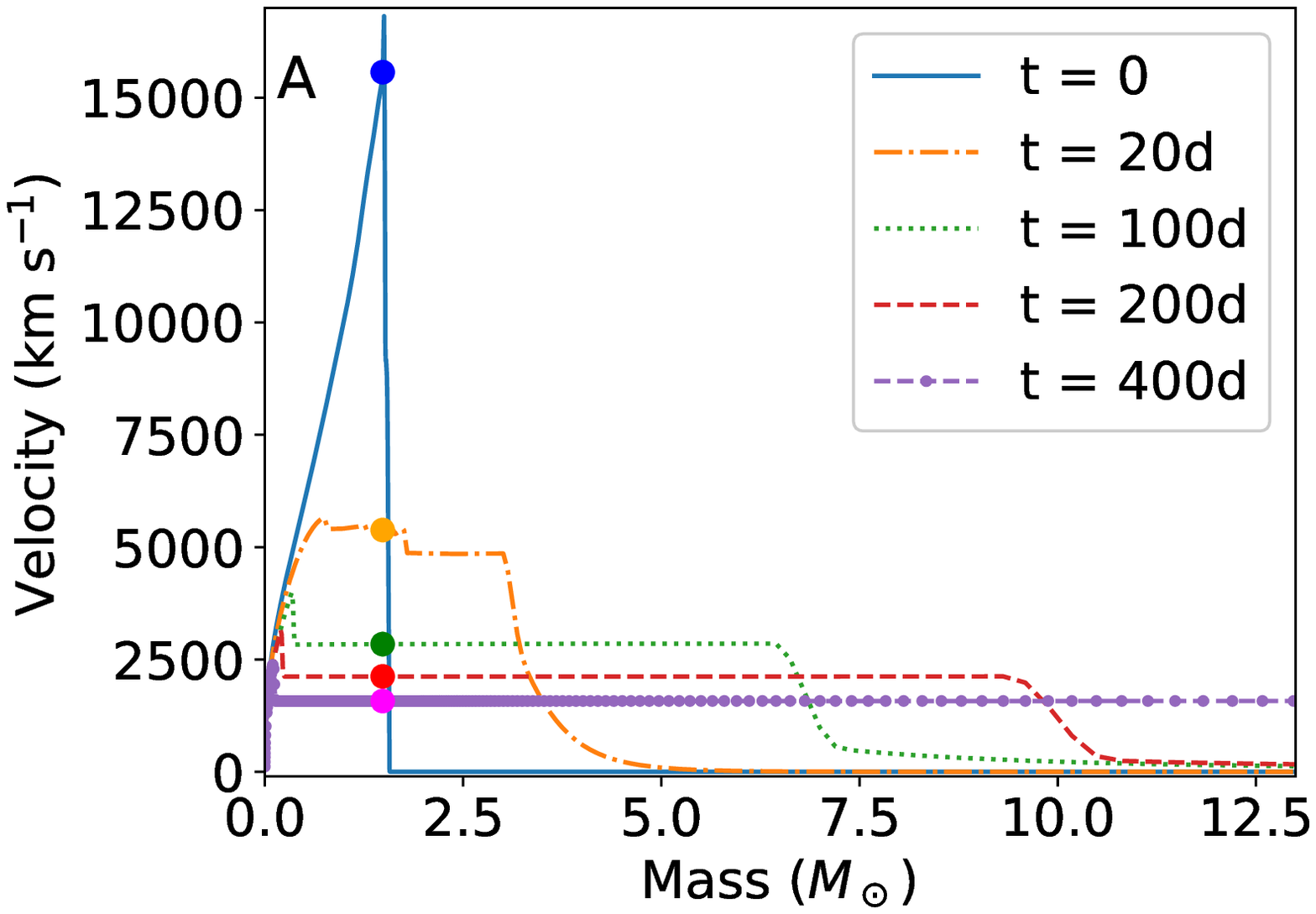}
\includegraphics[width=0.49\linewidth]{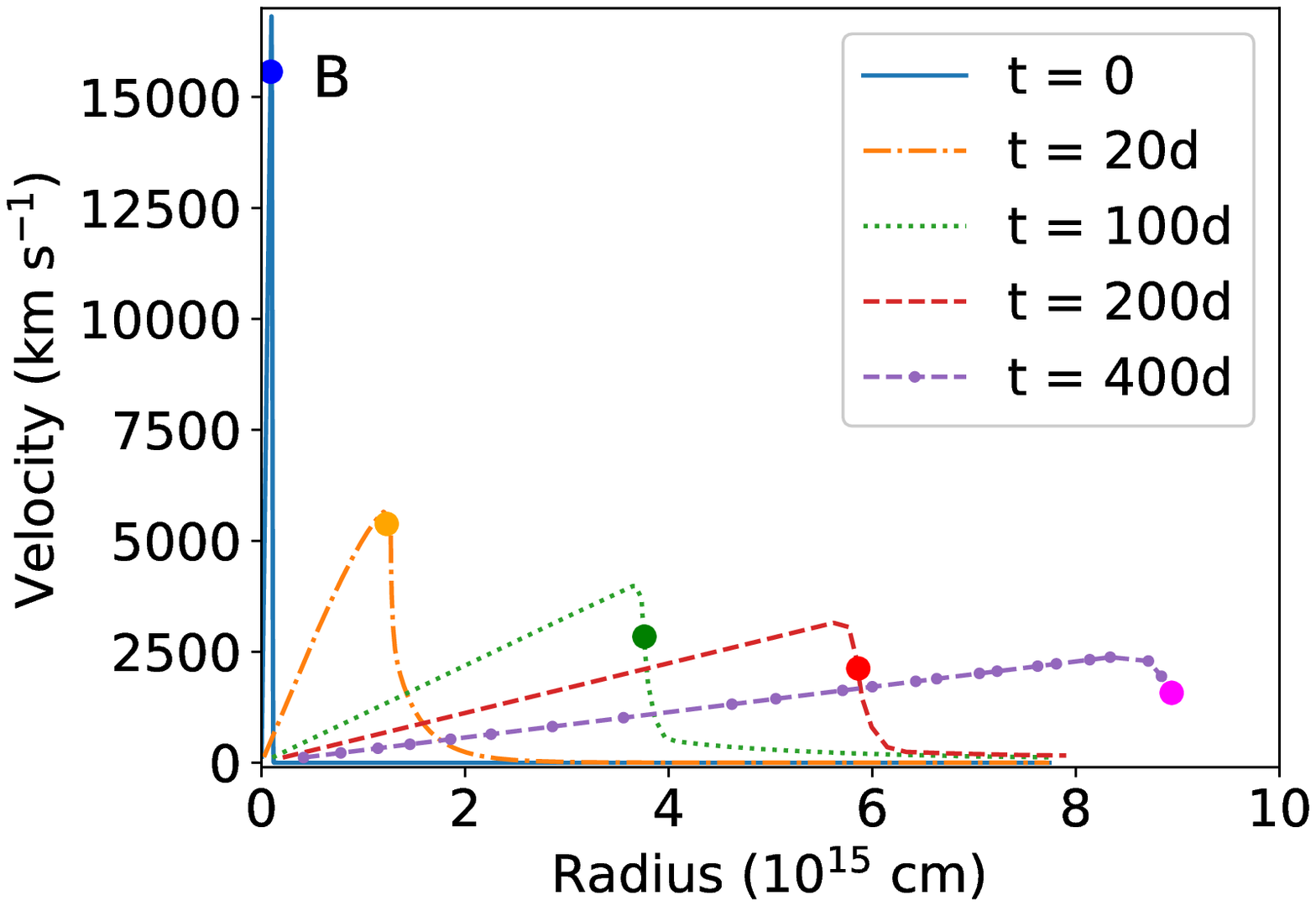}
\caption{\textbf{The dynamic evolution of the ejecta and CSM.} Velocity versus mass coordinate (panel A) and versus radius (panel B) for model $\dot{M}_{\rm CSM}=0.5~\rm M_{\odot}~\mbox{yr}^{-1}$, $R_{\rm CSM}=8\e{15}$ cm ($M_{\rm CSM}=13~\rm M_
\odot$). The interface between the ejecta and the CSM is marked by a circle.}
\label{fig:ejectaev}
\end{figure}

Figure \ref{fig:ejectaev} shows the dynamic evolution of the ejecta in the 13 \msun~model of Fig. \ref{fig:C}. The initial acceleration of the inner CSM gives velocities up towards 4000-5000 \kms. This region could give rise to broad hydrogen absorption features, as suggested by \cite{Smith2007} to be present in the spectra around peak, although it is difficult to unambigously identify such features. The similar supernova SN 2006tf \cite{Smith2008b} showed a spectral valley on the red side instead of the blue which means more mechanisms can be at play. At 400d essentially all ejecta+CSM have been collected into a shell with all mass moving with around 1500 \kms. 

We proceed to compare these CSM properties to previous modelling efforts of SN 2006gy where fundamentally different SN ejecta and explosion energies were explored.  \cite{Smith2007b} estimated a CSM mass of order 10 \msun~from diffusion time formulae, whereas \cite{Moriya2013} calculated a total of 15 models in a large 6-dimensional parameter space (ejecta mass, kinetic energy, shock velocity, CSM mass, CSM density profile, CSM extension). A difference to the Ia scenario is that the explosion energy in all these models is very large; $(5-50) \times 10^{51}$ erg. It is not straightforward to directly relate these simulations to the ones carried out here. Despite the large parameter space explored in \cite{Moriya2013}, the models show larger differences to SN 2006gy compared to the small set of physically constrained simulations carried out here. This may relate to the very large ejecta masses and energies in that grid, on the other hand the code used is more advanced than \textsc{\small SNEC}. Similarly, \cite{vanMarle2010} explored many models with $M_{\rm ejecta}=10-60$ \msun, none of them providing matches to SN 2006gy (their figure 25), although that was also not the main purpose of that paper.  \cite{Dessart2016} analyse in detail the dynamic and radiative properties of various interaction configurations, confirming with a more advanced method the basic properties of the dynamic evolution shown in Fig.  \ref{fig:ejectaev}.

In the pulsational pair-instability model of \cite{Woosley2007}, a 5 \msun~ejecta with $E=3\e{51}$ erg collides with a 25 \msun~CSM. These parameters are not too different from those of the Ia-CSM scenario. The presence of strong Fe~\textsc{\small I} lines means the pulsational pair instability scenario can likely be ruled out due to the lack of any nickel or iron in the ejecta. The light curve matches quite well as a CSM shell with $M_{\rm CSM} \gg M_{\rm ejecta}$ is located at the right place and extracts a large part of the kinetic energy of order $10^{51}$ erg, same as in the Ia-CSM scenario. This energy of the ejecta was set to four times higher than what is obtained in the pulsation simulations.

Given the relative simplicity of the light curve calculations and assumptions in our setup (e.g. continuous wind), the parameters derived for the CSM should be taken as indicative. For example, if the ejecta mass is smaller than 1.4 \msun~(possible for a sub-Chandrasekhar white dwarf explosion model, e.g., triggered by a detonation rather than deflagration), a smaller amount of CSM will be required to reach a similar level of the efficiency in converting the kinetic energy to the radiation. 

\begin{figure}
\centering
\includegraphics[width=1\linewidth]{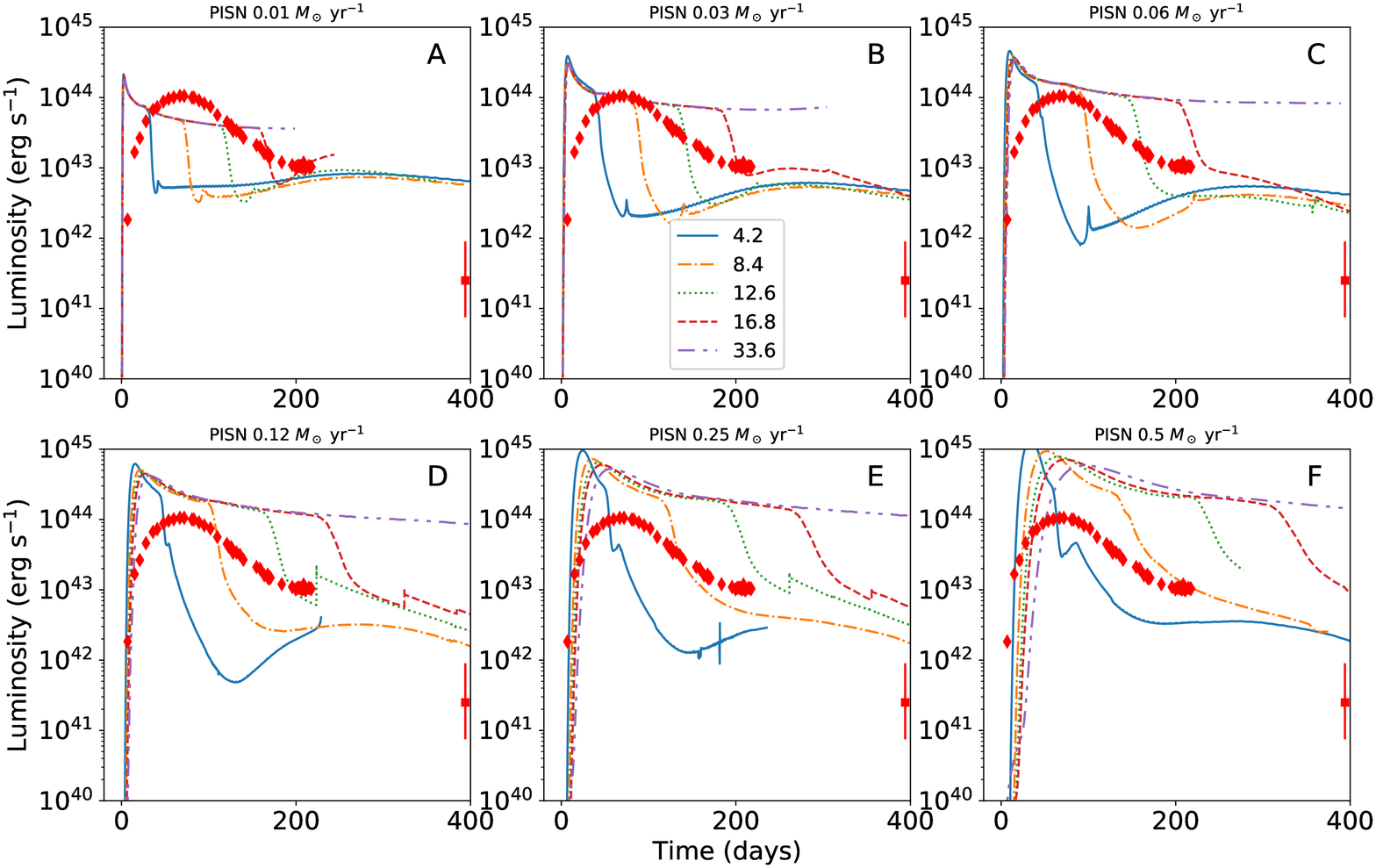}
\caption{\textbf{Model light curves for a He90 pair instability supernova interacting with a CSM, compared to SN 2006gy (red diamonds)}. Panels A-F show increasing values for the mass-loss-rate from 0.01 to 0.5 \msun~yr$^{-1}$.}
\label{fig:gridbig}
\end{figure}

\textbf{Pair-instability models.}
To investigate the scenario of pair instability supernovae, we calculated
also a \textsc{\small SNEC} grid where the He90 PISN model \cite{Heger2002}, the only model that can
explain the nebular spectrum, interacted with the same 2-parameter CSM.
Figure \ref{fig:gridbig} shows that the resulting light curves are in qualitative
disagreement with SN 2006gy. For the low mass loss rates, the diffusion time
is not sufficient to explain the characteristic time scale seen in SN
2006gy. For high mass loss rates, the diffusion time scale within the CSM
leads to a reasonable agreement of the observed light curve evolution time scale,
but the luminosity becomes much too high. These behaviors stem from much larger
kinetic energy in the PISN ejecta than the SN Ia model. For both limits, the late-time luminosity at 400d is also too high.

\begin{figure}
\centering
\includegraphics[width=1\linewidth]{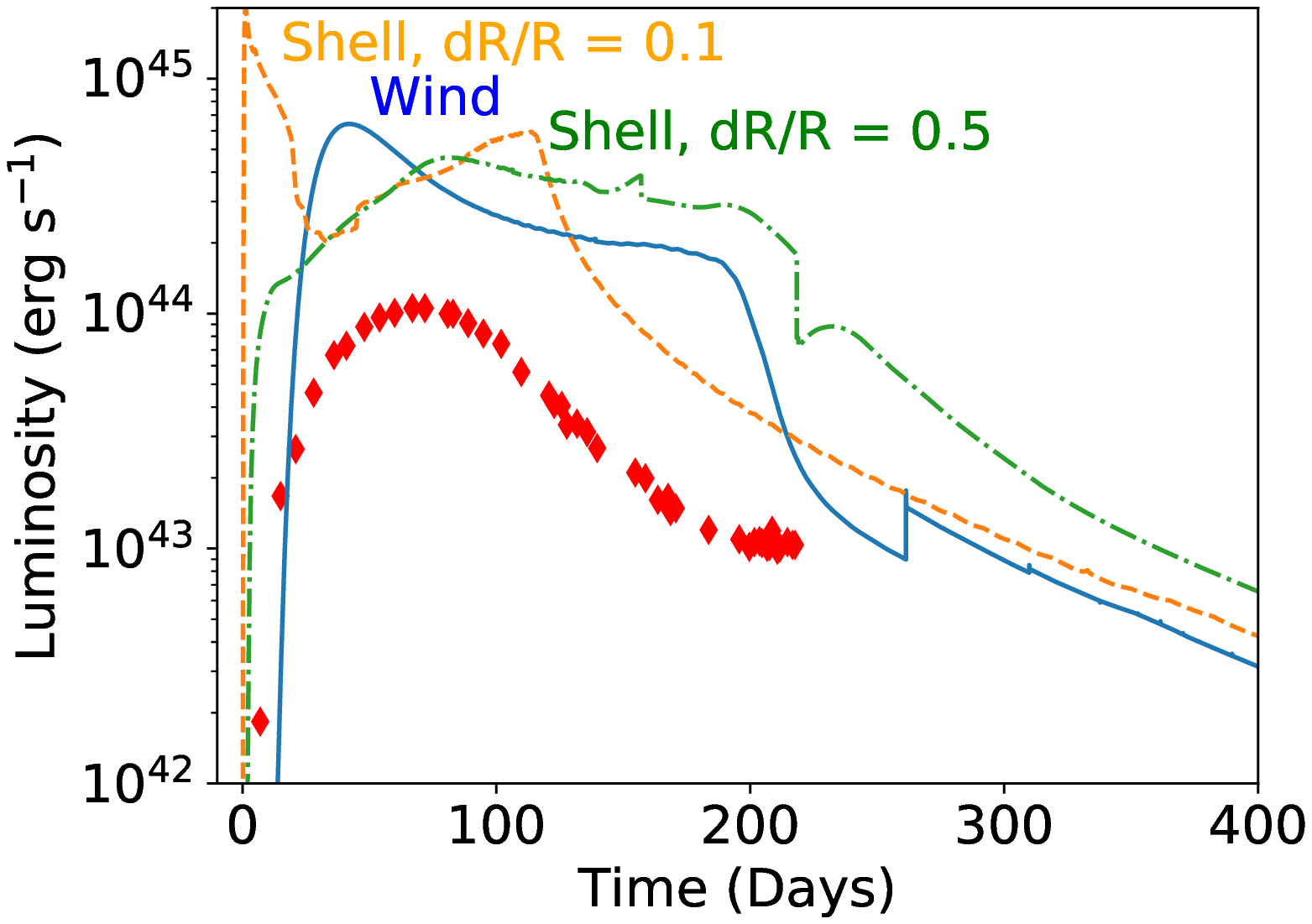}
\caption{\textbf{Comparison between wind and shell CSM distributions for PISN-interaction light curves}. Case $M_{\rm CSM}=10~\rm M_
\odot$, $R_{\rm CSM}=12.6\times 10^{15}$ cm. The SN 2006gy light curve derived here is plotted as red diamonds.}
\label{fig:PISNshells}
\end{figure}

We also explored whether other types of CSM morphology could improve the model fitting, but found no major improvements. Figure \ref{fig:PISNshells} compares the light curve from a wind distribution to shell distributions with two different thicknesses. While the light curve shapes can vary, the luminosity scale and the diffusion time-scale for the light curves are relatively similar, and none can match SN 2006gy. 

\subsection*{Supplementary text}
\subsubsection*{\underline{Energy budget.}} 
 \label{sec:energybudget}
Our analysis requires an accurate estimate of the total radiated energy of SN 2006gy. Typical Ia explosion models have $(1.3-1.4)\e{51}$ erg of kinetic energy \cite{Iwamoto1999}. Radiation hydrodynamic simulations, e.g. those in \cite{vanMarle2010} and those carried out here, show that a large fraction of this, $\sim70-95\%$ can be converted to radiation which allows for $\sim 10^{51}$ erg of radiated light. If the observed amount of light is clearly over this value, a Ia model can therefore be ruled out. 

The total radiated energy has previously been estimated by \cite{Smith2007} as $1.2\e{51}$ erg from $R$-band photometry,  assuming no bolometric correction ($M_{bol}=M_R$), \cite{Ofek2007} who obtained $1.1\e{51}$ erg using blackbody model fitting to $r$ and $i$ bands, but included only epochs up to 20d post-peak, and \cite{Smith2010} who estimated $2.5\e{51}$ erg by fitting blackbodies to spectra. These estimates assume similar host extinctions as here. 

However, rederiving the amount of radiated light by SN 2006gy using the full dataset including both spectra and photometry to estimate bolometric corrections rather than assumptions or blackbody fitting, results in a lower estimate for the radiated energy than in the original papers, in particular the high value obtained in \cite{Smith2010}. Our approach follows standard procedures for reconstructing the bolometric luminosity from a limited data set.  

Because the photometric data is far more complete in $R$-band than any other band, the method is based on relating $R$-band magnitudes to bolometric luminosities, using full spectral and photometric information at epochs where these are available to determine the mapping function. We first calculate the average $R$-band spectral flux level (erg s$^{-1}$ cm$^{-2}$ \AA$^{-1}$):
\begin{equation}
F_R = F_{ref,R} 10^{-m_R/2.5} \tag{S5}
\end{equation} 
where an apparent magnitude of zero corresponds to $F_{ref,R} = 2.2\e{-9}$ erg s$^{-1}$ cm$^{-2}$ \AA$^{-1}$ \cite{Bessel1979}.
We then seek a rescaling factor $s_{\rm tot}$ such that
\begin{equation}
F_R(t) \times \Delta \lambda_R \times s_{\rm tot}(t) = \int_{0}^{\infty} F_\lambda d\lambda = L_{bol}(t)/\left(4\pi d^2\right) \tag{S6}
\end{equation}
where $\Delta \lambda_R$ is a representative band width, $F_\lambda$ is the spectral flux, and $d$ is the distance (76.6 MPc). We adjust for calibration errors and slit losses by allowing a scalar calibration of each individual spectrum to fit the $R$-band photometry. Normally, one scales with a factor derived to minimize total error to all photometry. However, here the purpose is to relate $R$-band photometry to total luminosity, and for this the scaling has to occur to $R$-band (we are seeking the relative fluxes). The derived correction factors are all in the range 0.8-1.6, which is a typical range and shows that spectroscopic and photometric calibrations are consistent. We will find below that $s_{\rm tot}(t)$ is a slow-varying function over the time interval of interest and can be approximated as constant.

The arbitrarily defined band width $\Delta \lambda_R$ does not affect the results, but it allows us to visualize more easily the method by dividing the link between $F_R$ and the bolometric luminosity into an effective bandwidth $\Delta \lambda_R$ and a correction factor $s_{\rm tot}(t)$ to consider the flux outside this band. We choose $\Delta \lambda_R = 3000$ \AA~(the $R$-band filter covers $\sim$5500-9050 \AA), and remind again that the specific value has only pedagogic purpose. 

We use the well-sampled $R$-band light curve in \cite{Smith2007} as basis, which shows excellent agreement with other $R$-band data (e.g. \cite{Agnoletto2009}). As the optical $BVI$ photometry data is more plentiful than the NIR photometry, we in addition split the correction factor $s_{\rm tot}$ into three parts, first from $R$-band to optical (which we define as 3500-10000 \AA~based on the typical spectral coverage) called $s_{\rm opt}$, second from optical to quasi-bolometric (optical+NIR) called $s_{\rm NIR}$ (adding 10,000-25,000 \AA), and third to compensate for spectral ranges never observed ($<$3500 \AA~and $>25000$ \AA) called $s_{\rm UV+MIR}$, so $s_{\rm tot} = s_{\rm opt}\times s_{\rm NIR} \times s_{\rm UV+MIR}$. For the last factor, the combination of spectra and photometry show rapidly declining flux levels both below 3500 \AA~and beyond $K$-band so we assess corrections for these ranges to be minor for the time range considered. We allow a constant $s_{\rm UV+MIR}=1.1$ contribution by these bands based on by-eye inspection of the SEDs. As comparison, blackbodies between 3000-7000 K have  $s_{\rm UV+MIR} = 1.1-1.15$, and most of this is in the UV which is often blocked out in supernovae. 

The estimate of the total optical flux at epochs where spectra are available corresponds to scaling the spectra to $R$-band photometry and integrating. We now discuss the four epochs with multiband photometry from \cite{Agnoletto2009} (optical) and \cite{Miller2010} (NIR) and spectra available \cite{Smith2010}, which serve to determine the $s_{\rm tot}$ function.

\begin{figure*}
\centering
\includegraphics[width=0.45\linewidth]{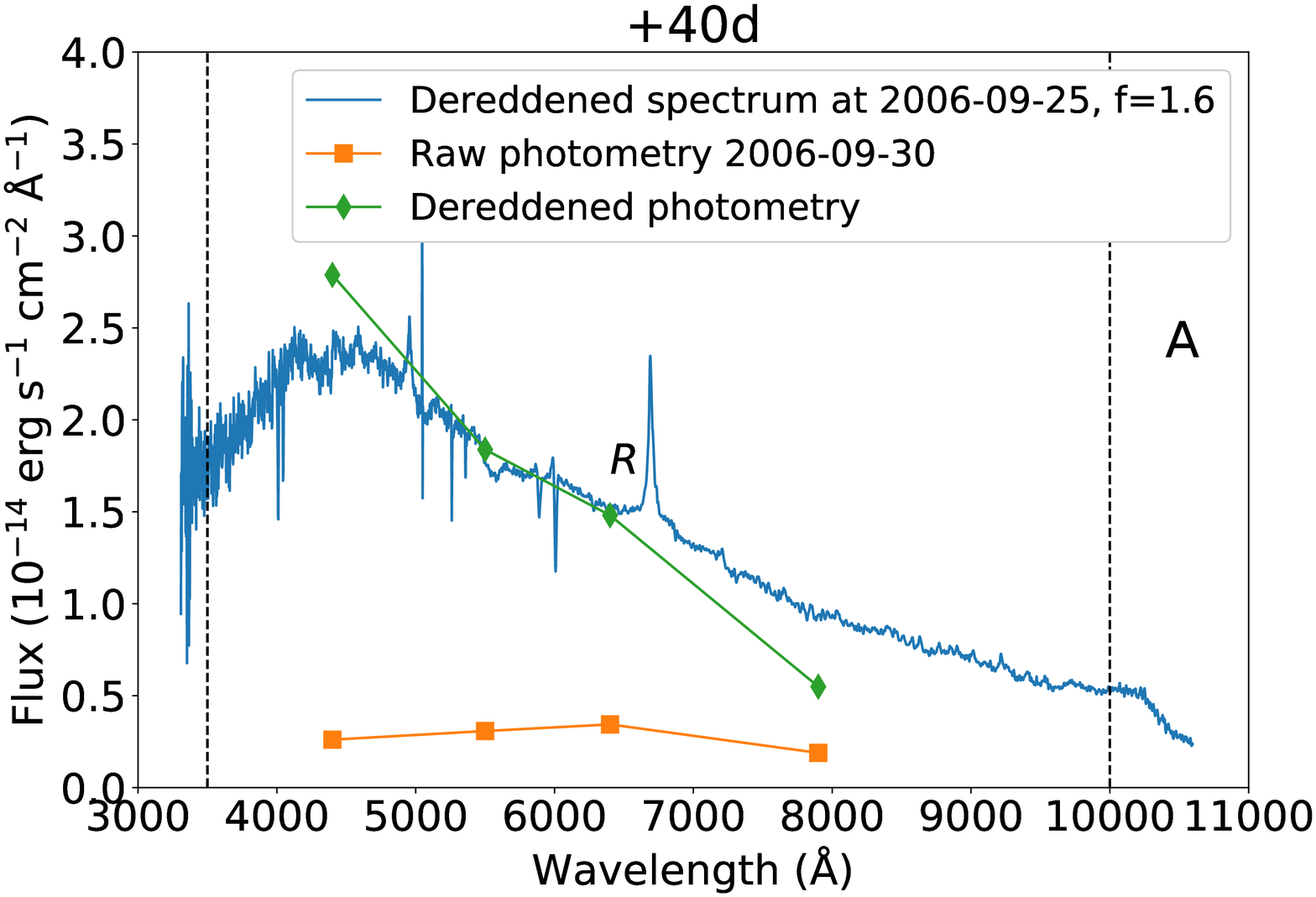}
\includegraphics[width=0.45\linewidth]{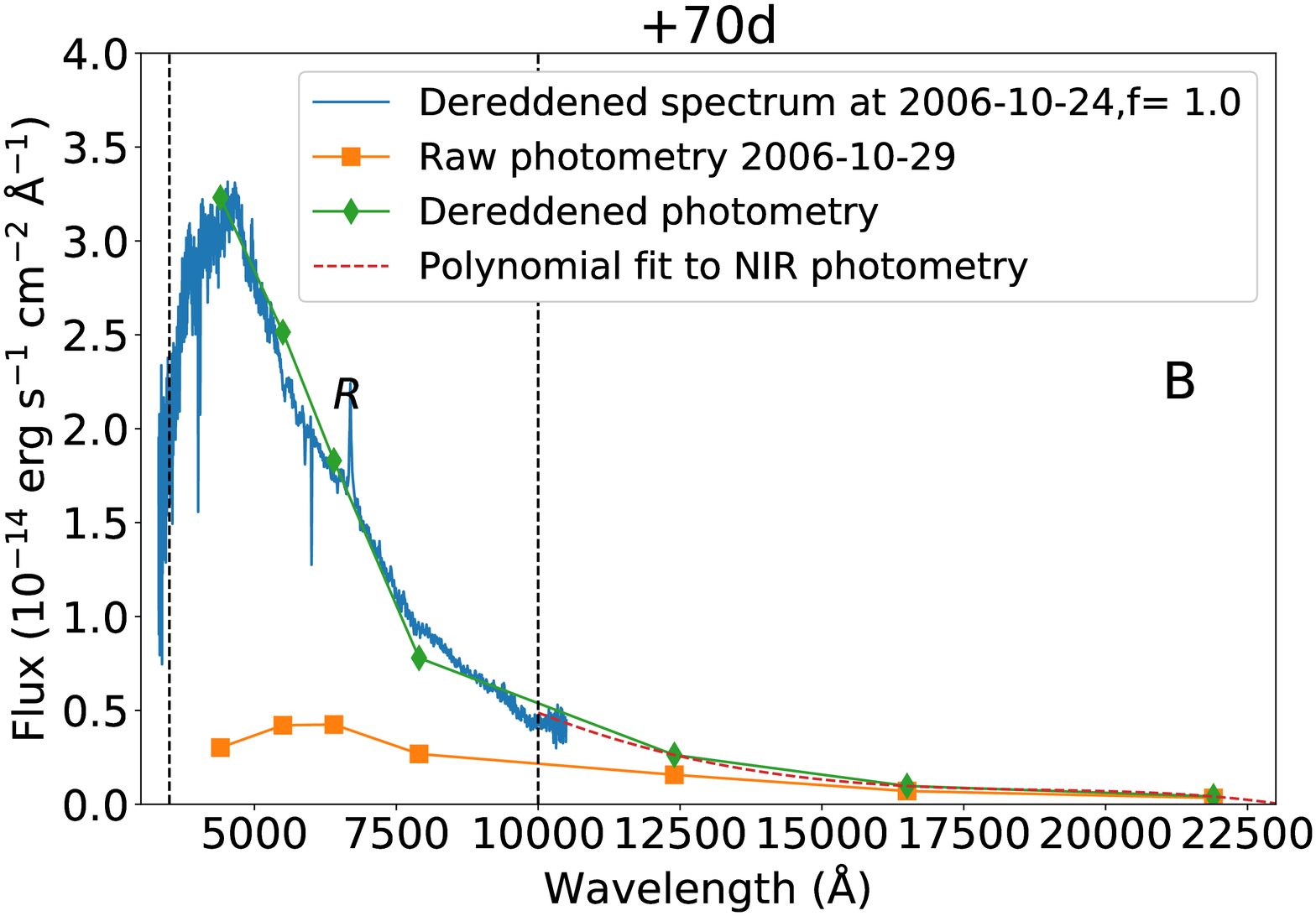}\\
\includegraphics[width=0.45\linewidth]{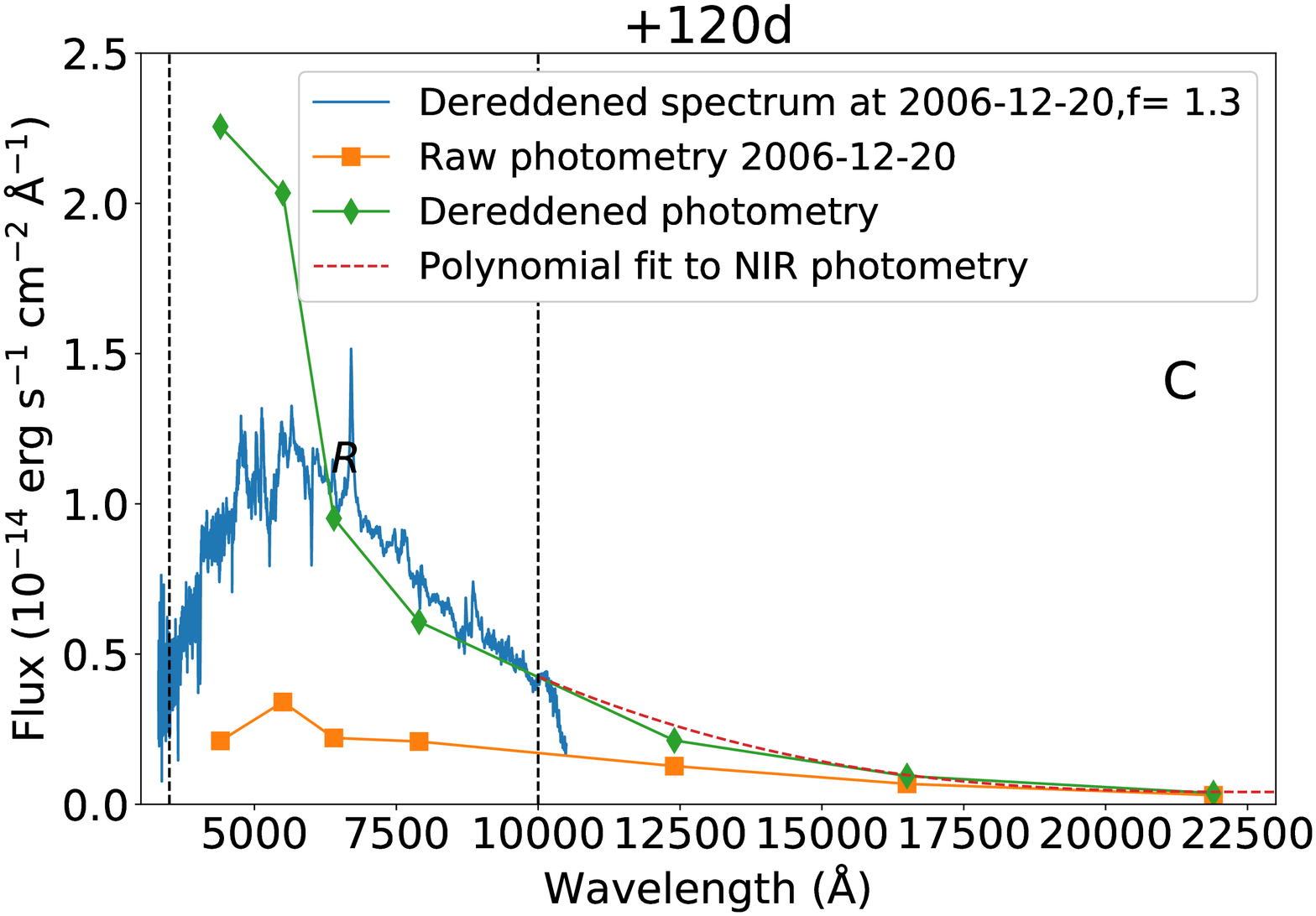}
\includegraphics[width=0.45\linewidth]{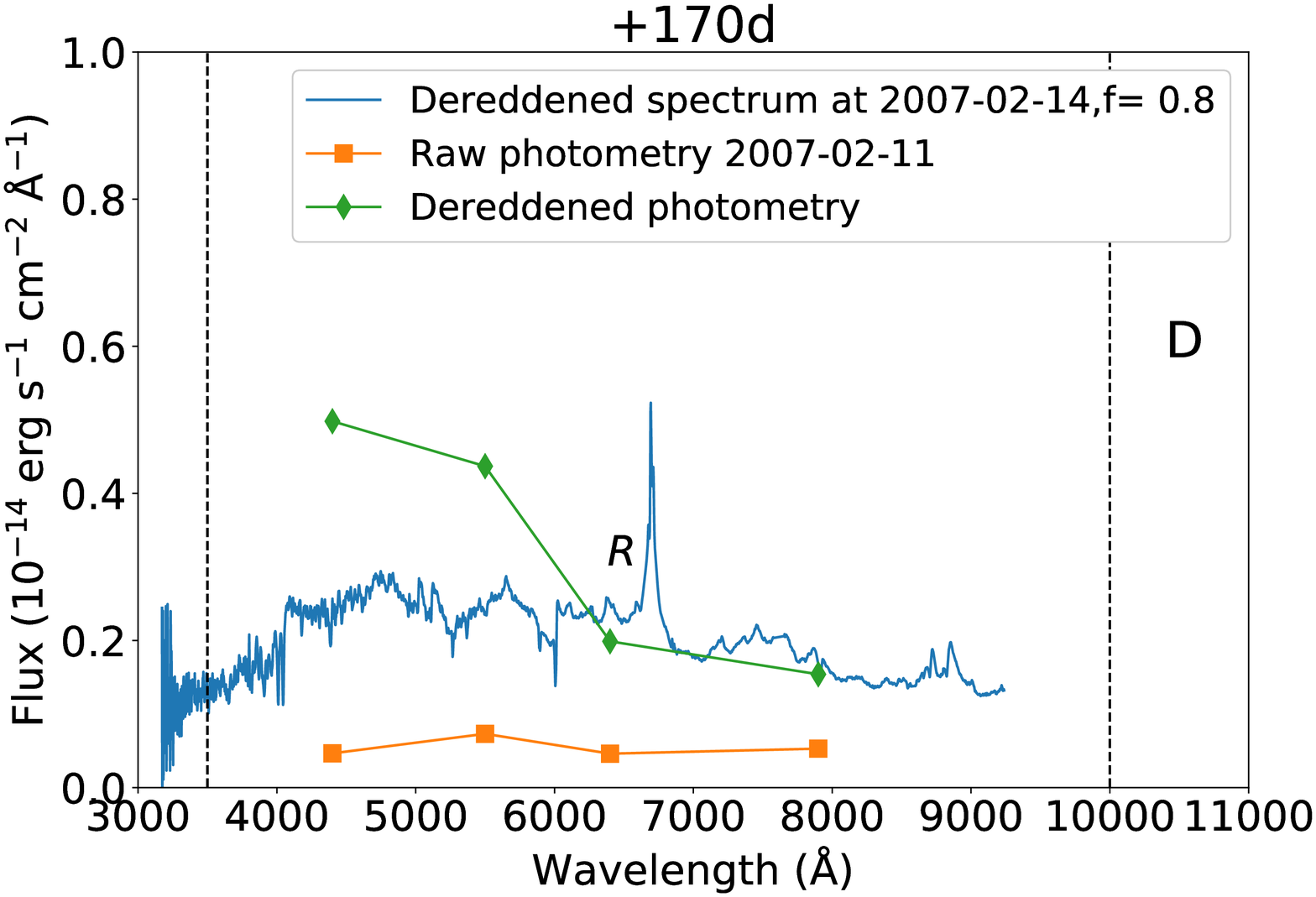}
\caption{\textbf{Spectra and photometry for SN 2006gy at the four epochs (covering the diffusion phase evolution) when both are available}. 
The epochs are 2006-09-30 (A), 2006-10-24 (B), 2006-12-20 (C) and 2007-02-14 (D). For each epoch, the dereddened spectrum (blue) is scaled to fit the (dereddened) $R$-band photometry, with scaling factors $f$ shown in the legend. The red dashed lines are fit functions to the NIR photometry. The panels cover different wavelength ranges depending on the available data for each epoch. The limits for our defined ``optical'' range (3500-10000 \AA) are plotted with vertical black dashed lines.} 
\label{fig:specpanel}
\end{figure*}

\paragraph{Epoch 1: 2006-09-25/30 (+40d)}
This is the first epoch with either a spectrum or multi-band photometry available, so the first epoch for which we can estimate a correction factor.
We follow \cite{Smith2007} and estimate an explosion epoch of August 20 2006 (Modified Julian Date MJD 53967), six days before the first non-detection point.
Figure \ref{fig:specpanel} (upper left) shows the scaled extinction-corrected spectrum and photometry, which are in good agreement. The flux correction factor from $R$-band to optical is $s_{\rm opt}=1.92$. There is no NIR data.

\paragraph{Epoch 2: 2006-10-24/29 (+70d)}
At this epoch (close to peak) there is a spectrum, $BVRI$ photometry, and NIR photometry. Figure \ref{fig:specpanel} (upper right)  shows good agreement between scaled spectra and photometry. The $R$-band to optical correction factor is $s_{\rm opt}=1.82$. The NIR correction factor (from optical to optical+NIR) is $s_{\rm NIR}=1.16$, obtained by fitting a 4th order polynomial to the NIR photometry and integrating between 10,000-25,000 \AA.

\paragraph{Epoch 3: 2006-12-19/20 (+120d)}
At this epoch there is a spectrum, $BVRI$ photometry, and NIR photometry. Figure \ref{fig:specpanel} (bottom left) shows there is some disagreement between the SEDs obtained by spectrum and photometry. The photometry shows a bluer SED compared to the spectrum. The $R$-band to optical correction factor is $s_{\rm opt}=1.76$ (2.55 for photometry). An independently observed spectrum at this epoch shown in figure 5 in \cite{Agnoletto2009} is very similar to the spectrum shown here, and spectra at other nearby epochs are similar in \cite{Smith2007}. This suggests that it is the $B$ and $V$ photometry that may be too high. However, below we assess the impact of both assumptions. The NIR correction factor (from optical to optical+NIR) is $s_{\rm NIR}=1.35$.

\paragraph{Epoch 4: 2006-02-11 (+170d).}
At this epoch there is a spectrum and $BVRI$ photometry, but no NIR photometry. Figure \ref{fig:specpanel} (bottom right) shows there is some disagreement between the spectral and photometric colors, similar to the previous epoch. The $R$-band to optical correction factor is $s_{\rm opt}=1.99$ (2.98 for photometry).

\paragraph{Correction factor evolution.}
At the two epochs where there is no NIR data we assumed the same correction factor as at the nearest measured epoch.
Table \ref{table:corrfacs} and Figure \ref{fig:f0} shows that the total correction factor $s_{\rm tot}$ lies close to 2.6 at all times. 
This means that the bolometric light curve has a similar morphology as the $R$-band light curve, and that we can to good accuracy integrate the $R$-band total emission ($3.3\e{50}$ erg for $\Delta \lambda_R=3000$ \AA) and just multiply by 2.6 (error will be of order 10\%). This gives us a final estimated bolometric energy of $8.6\e{50}$ erg.
Should the photometry for the last two epochs be more accurate than the spectra, the correction factors at those epochs rise by 44\% (2006-12-20) and 49\% (2007-02-14), respectively. As only the post-peak part of the light curve is affected, the total energy increases by 25\% to $1.1\e{51}$ erg. This difference is smaller than the uncertainty in the extinction.

\begin{table}
\caption{\textbf{Correction factors to convert $R$-band luminosity (assuming 3000 \AA~band width) to bolometric luminosity.}}
\label{table:corrfacs}
\begin{center}
\begin{tabular}{ccccccc}
\hline
Date           & Phase  &  $s_{\rm opt}$ & $s_{\rm NIR}$ & $s_{\rm UV+MIR}$ & $s_{\rm tot} = s_{\rm opt}\times s_{\rm NIR} \times s_{\rm UV+MIR}$  \\
\hline
2006-09-25 & +40d    & 1.92          & n/a              &   1.1            & 2.45  \\
2006-10-24 & +70d   & 1.82          & 1.16            &    1.1           & 2.32 \\
2006-12-20 & +120d   & 1.76          & 1.35            &    1.1           & 2.61  \\
2007-02-11 & +170d & 1.99         & n/a              &    1.1           & 2.95 \\
\hline
\end{tabular}
\end{center}
\end{table}

\begin{figure}
\includegraphics[width=1\linewidth]{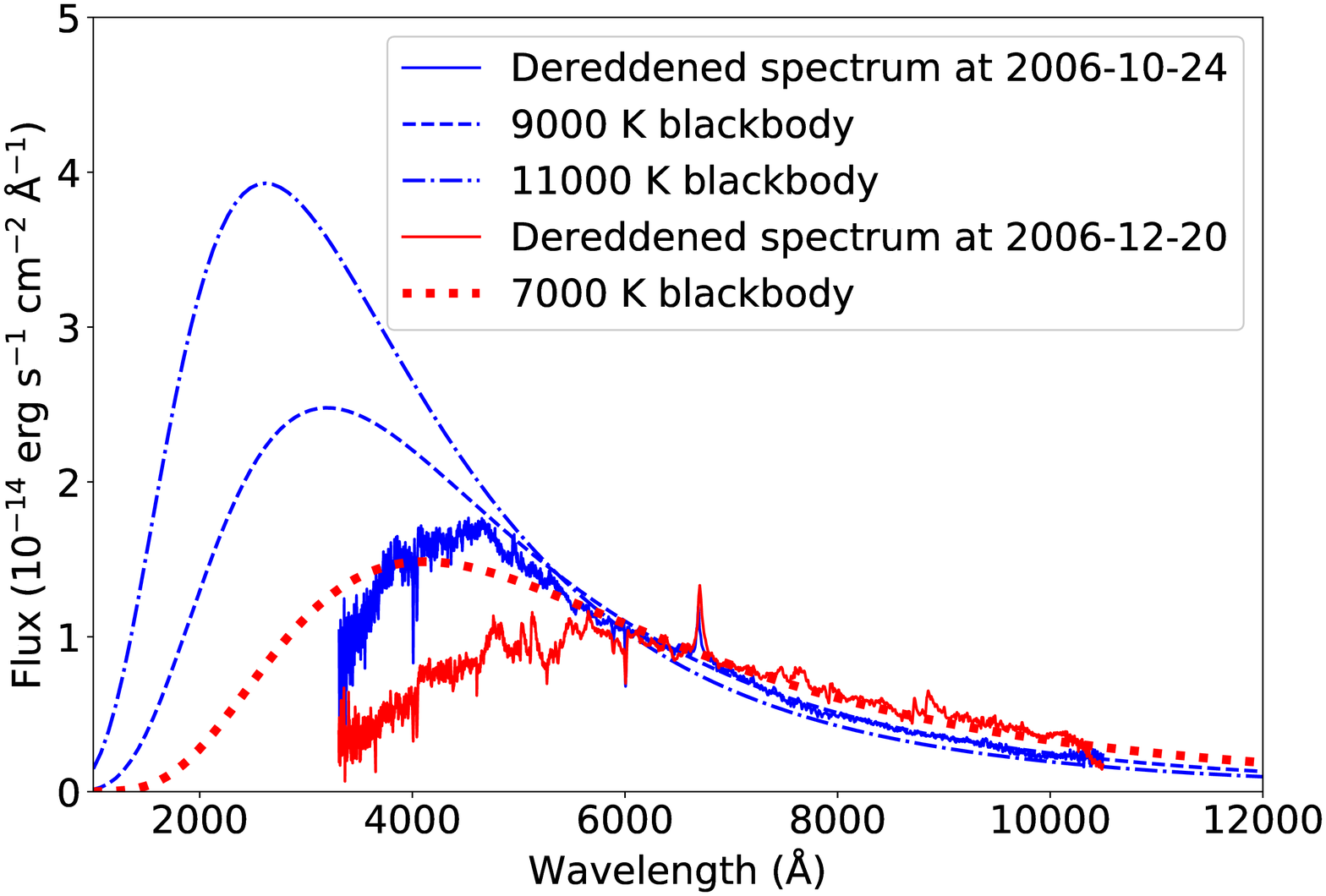}
\caption{\textbf{Comparison between dereddened spectra of SN 2006gy (observations by \cite{Smith2010}, here rescaled to overlap at $R$-band centre) and blackbodies}. The use of blackbodies can lead to a significant overestimate of the bolometric luminosity if the SN shows signatures of line blocking, as SN 2006gy does. These temperatures are those fitted in \cite{Smith2010} in their derivation of an emitted energy of $(2-3)\times 10^{51}$ erg.}
\label{fig:vega}
\end{figure}

\begin{figure}
\includegraphics[width=1\linewidth]{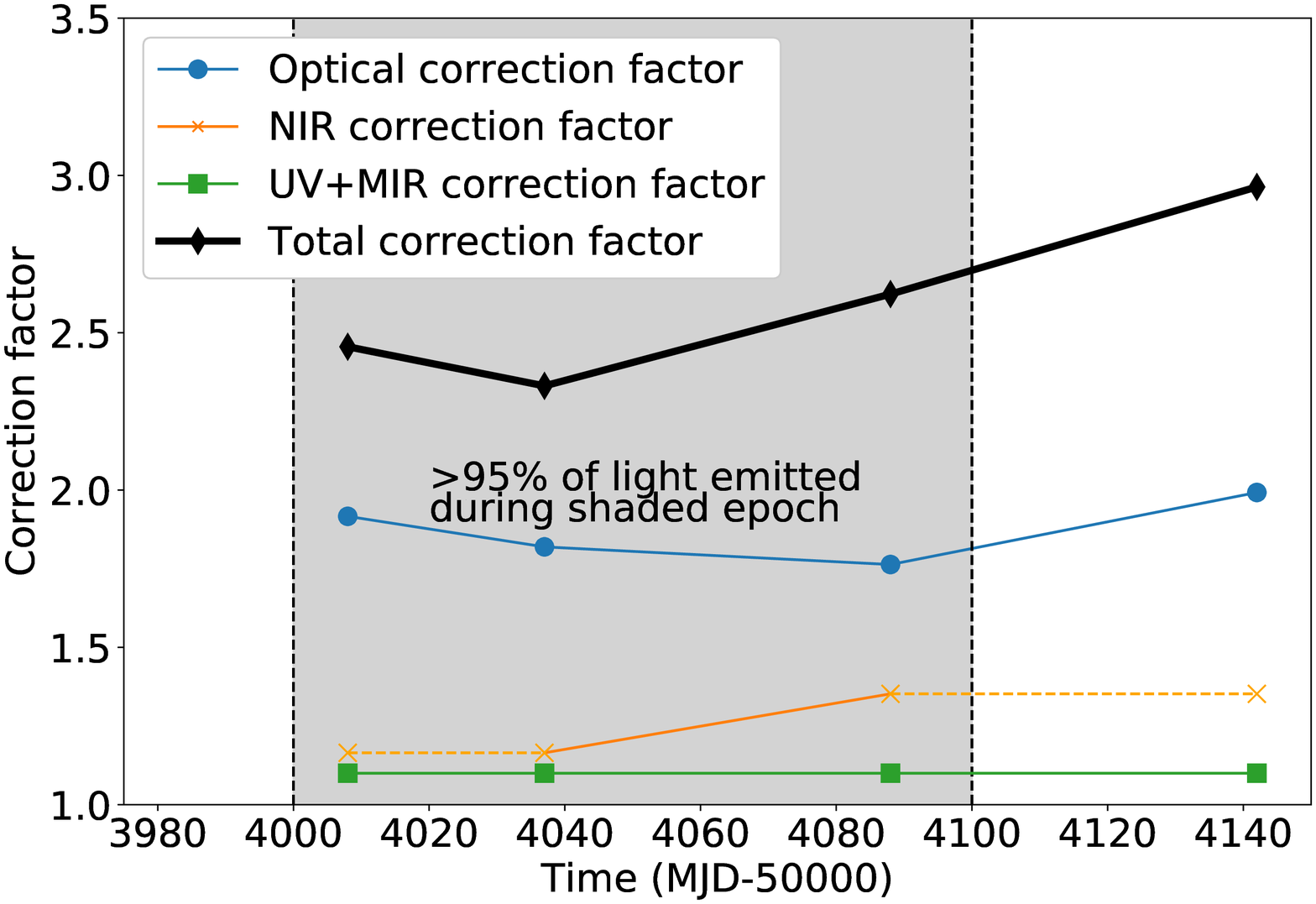}
\caption{\textbf{Calculated correction factors $s_{\rm opt}$ (blue), $s_{\rm NIR}$ (orange), $s_{\rm UV+MIR}$ (red) and their product $s_{\rm tot}$ (black)}. Over the 100 days over which the supernova emits $>$95\% of its light (shaded gray), the $s_{\rm tot}$ value is always close to 2.6.}
\label{fig:f0}
\end{figure}

The \cite{Smith2010} result of $(2-3)\e{51}$ erg has in particular widely been taken to imply that a Ia-CSM scenario can be ruled out as the energy exceeds by a factor 2-3 the maximum output of such models. However, our lower estimate gives a total emitted energy close to the expected value for Ia-CSM models in the limit $M_{\rm CSM} \gg 1.4$ \msun. It is therefore relevant to consider the source of the difference in more detail. The method of \cite{Smith2010} is described in their section 3.2. Blackbody temperatures are first estimated (their figure 7, middle panel), and then observed $R$-band photometry is used to set the absolute scale of these blackbody curves at each epoch. The SN 2006gy spectra show however quite strong deviations from blackbody shapes, in particular being dimmer at blue wavelengths than blackbody extrapolations from red bands predict (Fig. \ref{fig:vega} here and figure 4 in \cite{Smith2010}, see also \cite{Fransson2014} and \cite{Dessart2015} for UV properties in models and observations of other IIn SNe). This approach may therefore give an overestimate of the total luminosity.

We also discuss the report from \cite{Ofek2007} of a radiated energy of at least $1.1\e{51}$ erg. These authors fit blackbodies to $r$ and $i$-band photometry, and again the strong deviation from blackbody shapes for SN 2006gy may lead to an overestimate.
In addition, the temperature errors are quite large when fitting only the Jeans tail of the blackbody and this maps to large luminosity errors through the $T^4$ dependency.

In conclusion, we assess that the differences to earlier methodologies can be understood, and that the value derived here is reasonable. We use SEDs to derive correction factors rather than fitting blackbodies, and do not rely on assumptions about the SED as for the method of assuming zero bolometric correction. We use average values for reported estimates of distance and extinction, and estimates for the missing flux in UV and MIR.

This factor 2-3 reduction of the total radiated energy is important because it brings Ia-CSM models back into contention, with the best estimate for the total radiated energy to the level expected for a Ia SN in the strong interaction limit ($M_{\rm CSM} \gg M_{\rm SN})$. The quite large uncertainty in extinction for SN 2006gy means an exact value cannot be derived, but combining both the downward revision here from SED methodology and extinction uncertainty, one cannot rule out a Ia-CSM scenario on energy grounds, and in fact our best estimate is at the predicted value in this scenario. Even if the radiated energy is higher, super-Chandrasekar Ia explosions (as necessarily made by mergers) can theoretically reach energies of around $2\e{51}$ erg \cite{Khokhlov1993}.\\

\noindent \textbf{Kinetic energy and energy from late echoes.}\\
The initial kinetic energy of the supernova ($\sim 1.3\e{51}$ erg in the standard explosion scenario, \cite{Iwamoto1999}) has thus been divided into radiation and kinetic energy of the SN+CSM mass, and with time the first quantity increases at the expense of the second. By +394d, the SN+CSM ejecta moves with around 1500 \kms (inferred from the line profiles), which gives a kinetic energy of $\sim 1/2 \times 15 \times 2 \times 10^{33}
 \times \left(1500\times 10^5\right)^2 \sim 3\e{50}$ erg.

The radiation observed at yet later times is dominated by echoes. The analysis of this is difficult and uncertain as it lies mostly in the MIR with very sparse observational coverage. 
An echo energy cannot straightforwardly be linked to a total radiated energy budget as that link depends on the morphology of the dust distribution (only for a spherically symmetric case could the echo be added up to the budget). For such an assumption,
\cite{Fox2015} obtain estimates between $3\e{49}-4\e{50}$ erg depending on the model, and for this range the echo energy would become a moderate correction to the total value (3-40\%).  

\subsubsection*{The very late evolution.}
In \cite{Miller2010} (see also \cite{Smith2008}) late-time $K'$-band magnitudes of about 15 at +400d are reported. With the $K'$ zero-point at 25.92 that corresponds to $4.3\times 10^{-17}$ erg s$^{-1}$ cm$^{-2}$ \AA$^{-1}$, or $1.0\e{41}$ erg s$^{-1}$ for a bandwith of 3500 \AA. The IR SED (their figure 7) shows that this is the beginning of a cool, secondary SED component likely corresponding to the thermalized part of a dust echo. With much less flux in the $H$ band (2 magnitudes), not much extra direct emission from the SN itself (rather than the dust echo) is present in the NIR at +400d compared to the optical, and the dust temperature is inferred to be cool, $<$ 1000 K.  
The interpretation of the $K'$ emission with a dust echo requires 1-2\% of the emitted light from the diffusion peak to be reradiated, which \cite{Miller2010} find to be similar as in other IIn supernovae. Finally, note that this dust is located much further away than the CSM that causes the interaction light curve, and it is unclear if it has any direct relation the the progenitor system \cite{Miller2010}.

Important for the data reduction and echo removal is also the +810 optical photometry reported by \cite{Miller2010}. While the IR emission is interpreted as thermal radiation from the dust, the optical spectrum is interpreted as due to scattered light either from this same dust or from another more distant dust region. The main argument for this is that spectrum at +810 day is bluer than at earlier epochs. This is consistent with the strong wavelength-dependence of dust scattering, and is not reproduced by most models for SN emission itself which predict cooling and reddening with time. 
An exception would be multiple shell interactions as in certain pulsational PISN models \cite{Woosley2017}. One would then, however, also expect rebrightening of the light curve which is not seen in SN 2006gy. Taken together with the fact that the SED looks as expected in the echo scenario (the peak SED filtered though a $\lambda^{(-1)-(-2)}$ power law), we find this the most convincing interpretation.
The inferred (uncorrected) optical luminosity of the echo at +810d is $3\times 10^{40}$ erg s$^{-1}$. If the optical echo has had same strength (models predict constant or slow declining echo strengths \cite{Chevalier1986}) it is therefore likely that it influences the +394d spectrum at a similar level, and an assumption that it has the same strength fits the +394d quasi-continuum well.  
The +810d flux in the red band containing the Fe~\textsc{\small I} lines at +394d is much below that of +394d, and those Fe~\textsc{\small I} lines are also not seen in any earlier spectra, which eliminates the possibility that they are echoed at +394d. 

\subsubsection*{\underline{Further tests of Fe masses}}
With the parameterized modelling we have shown that, assuming $f=0.1-1$, the Fe~\textsc{\small I} lines cannot emerge at all under any physical conditions unless $M(\mbox{Fe})$ $>$ 0.05 \msun, and only in the right relation to Fe~\textsc{\small II} for $M(\mbox{Fe})$ $>$ 0.3 \msun. A filling factor of $f=0.1$ corresponds to a shell thickness of $dV/V=3\%$ for V=1500 \kms~and is a plausible limit to the clumping/compression. The multi-zone W7+CSM model confirms that a model with a \ni~mass of 0.5 \msun~works well when all conditions are self-consistently calculated.

We consider the Fe mass limit by further numeric experiments. The parameterized models have the advantage that the constraints they give are independent of the specific details of the ejecta composition and powering; they explore any possible combinations of temperature and ionization and therefore give results holding for any composition. Unless the basic assumptions are violated, they should not give any false negatives (rule-out of solutions that are in fact viable). As such physical property limits derived with them should be conservative. They may however, give false positives - combinations of temperature and ionization that cannot be achieved by any realistic ejecta composition or powering setup for the given mass and filling factor. Modelling from the other direction - self-consistent models with a specific structure - can cast some light on distinguishing such false positives. However, here one deals with a basically infinite parameter space for the ejecta details. Thus, we can never be sure such modelling has delineated all the allowed solution space.

\begin{figure*}
\centering
\includegraphics[width=0.8\linewidth]{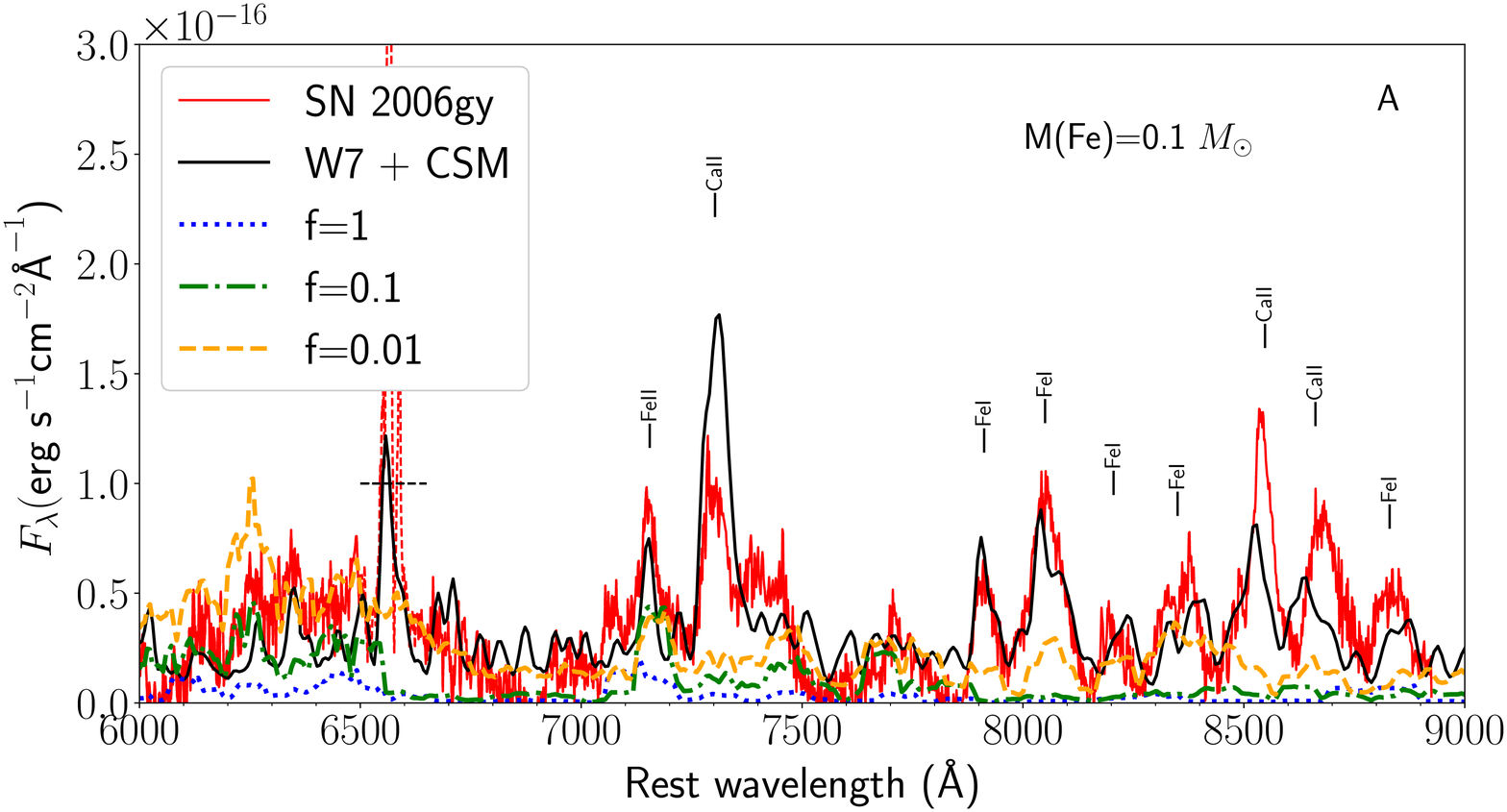}
\includegraphics[width=0.8\linewidth]{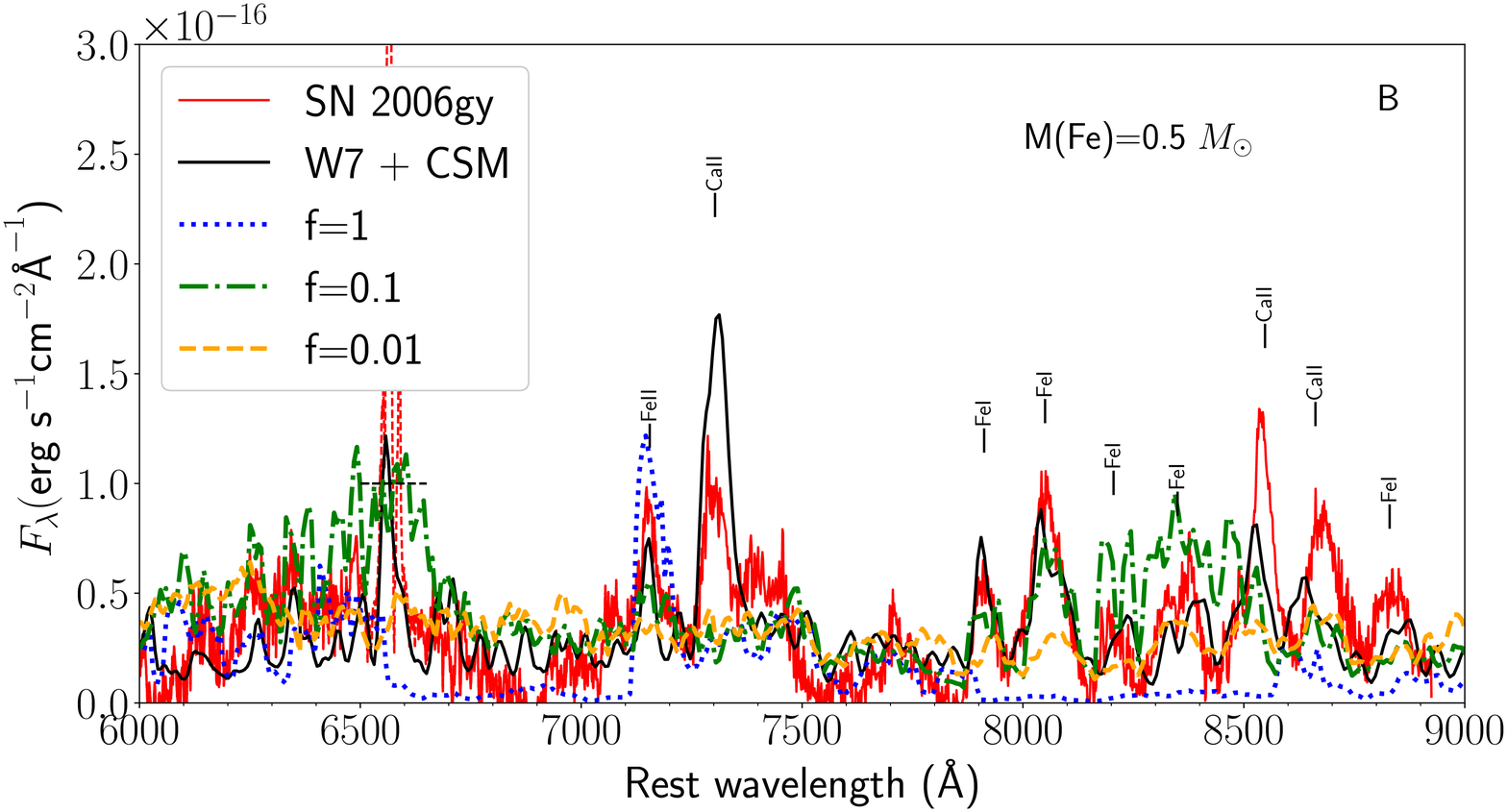}
\includegraphics[width=0.8\linewidth]{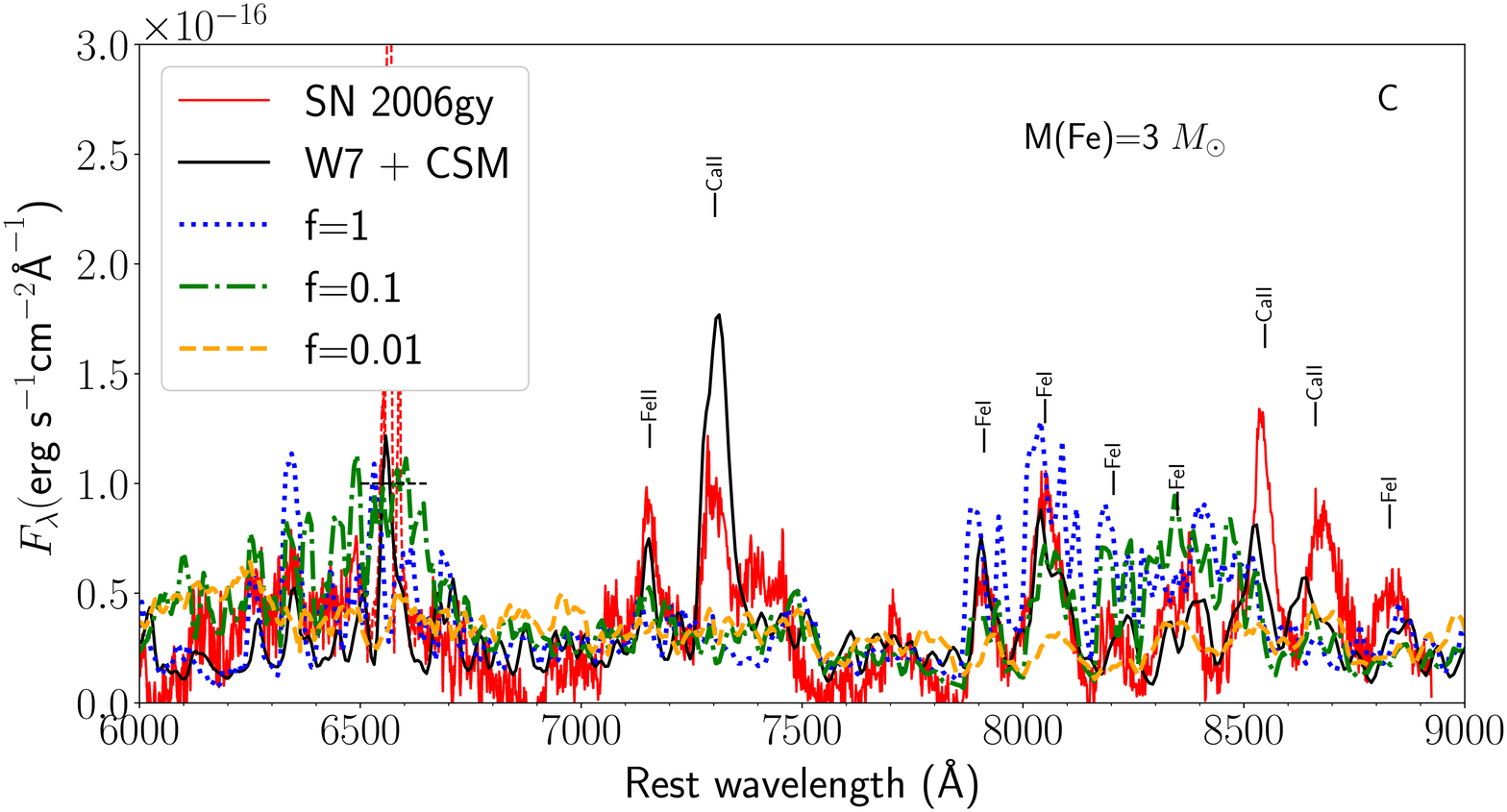}
\caption{\textbf{Multi-zone self-consistent \textsc{\small SUMO} models of Fe shells with different masses and filling factors}. The Fe~\textsc{\small I} lines
emerge for large Fe masses of 0.5 \msun~(panel B) and 3 \msun (panel C) but not for low ones (0.1 \msun, panel A), even for $f=0.01$.}
\label{fig:varyFe}
\end{figure*}

We use further modelling to anchor the lower mass limit (0.3 \msun) that was derived above. One concern is whether a lower filling factor than $f=0.1$ (although not easily obtainable) could allow for a lower Fe mass. To address this, we calculated additional self-consistent \textsc{\small SUMO} models of Fe shells with $M(\mbox{Fe})$=0.1 \msun~and $f=0.01,0.1,1$. All models have a fixed energy input of $2.5\e{41}$ erg s$^{-1}$ (the estimated bolometric luminosity at 400d). Resulting spectra are shown in Fig. \ref{fig:varyFe}. As expected, the $f=0.1$ and $f=1$ models fail; there is no Fe~\textsc{\small I} emission. At $f=0.01$ (and lower) the Fe~\textsc{\small I} lines also do not emerge distinctly. While the ejecta becomes more neutral at higher compression, the higher densities also leads to more continuum-like spectra. 

The middle and bottom panels show that at larger Fe masses the Fe~\textsc{\small I} lines do emerge, at least for some filling factors, and further supports the positive solution spaces in the single-zone parameterized modelling for such masses. With this simple model setup (just a constant density shell of pure Fe) the match to the lines is notably worse than the W7+CSM ejecta, which can be taken as further merits of that specific model.

\subsubsection*{\underline{Common envelope scenarios}}

In addition to the references in the main text, further common envelope simulations of relevance for the Ia-CSM scenario for SN 2006gy can be found in \cite{Sandquist1998} and \cite{Yorke1995}.

One possibility is an AGB (Asymptotic Giant Branch) + WD merger, which would involve two degenerate cores which has been shown to be able to produce a Type Ia explosion \cite{Pakmor2012}. Our best fitting CSM mass is about a factor two larger than a standard AGB star could provide ($\sim$6 \msun), but due to the simplifications in the modelling a 6 \msun~CSM scenario is not necessarily excluded. 

Another possibility is a non-degenerate core of the companion star, e.g. the He core of a supergiant. Accretion and possible explosion can in this case happen at a larger separation (about $1~\rm R_\odot$), and the available envelope mass is larger, up to 20 $\rm M_\odot$. 
Inpiral of a neutron star into the helium core of a blue supergiant (BSG) happens on a time-scale of 10y in simulations, with associated ejection of a large fraction of the envelope \cite{Terman1995}, and similar dynamics would be expected if the neutron star was replaced by a WD. The formation of BSG/RSG + WD binaries are predicted by binary population synthesis codes \cite{Tutukov1993, Tauris2000,Willems2004}, and candidate systems have been observed \cite{vanKerkwijk1999}. The uncertainty in this scenario is whether a He core + WD merger is a robust path to explosion, although 
the existence of the He detonation-triggered SN Ia has been recently suggested observationally \cite{Jiang2017}. 
 
\end{document}

%% file: science_macros.tex
\def\jnl@style{\it}
\def\aaref@jnl#1{{\jnl@style#1}}
\def\aaref@jnl#1{{\jnl@style#1}}

\def\aj{\aaref@jnl{Astron. J.}}                   % Astronomical Journal
\def\araa{\aaref@jnl{Annu. Rev. Astron. Astrophys.}}             % Annual Review of Astron and Astrophys
\def\apj{\aaref@jnl{Astrophys. J.}}                 % Astrophysical Journal
\def\apjl{\aaref@jnl{Astrophys. J.}}                % Astrophysical Journal, Letters
\def\apjs{\aaref@jnl{Astrophys. J. Suppl. Ser.}}               % Astrophysical Journal, Supplement
\def\ao{\aaref@jnl{Appl.~Opt.}}           % Applied Optics
\def\apss{\aaref@jnl{Ap\&SS}}             % Astrophysics and Space Science
\def\aap{\aaref@jnl{Astron. Astrophys.}}                % Astronomy and Astrophysics
\def\aapr{\aaref@jnl{A\&A~Rev.}}          % Astronomy and Astrophysics Reviews
\def\aaps{\aaref@jnl{A\&AS}}              % Astronomy and Astrophysics, Supplement
\def\azh{\aaref@jnl{AZh}}                 % Astronomicheskii Zhurnal
\def\baas{\aaref@jnl{BAAS}}               % Bulletin of the AAS
\def\jrasc{\aaref@jnl{JRASC}}             % Journal of the RAS of Canada
\def\memras{\aaref@jnl{MmRAS}}            % Memoirs of the RAS
\def\mnras{\aaref@jnl{Mon. Not. R. Astron. Soc.}}             % Monthly Notices of the RAS
\def\pra{\aaref@jnl{Phys.~Rev.~A}}        % Physical Review A: General Physics
\def\prb{\aaref@jnl{Phys.~Rev.~B}}        % Physical Review B: Solid State
\def\prc{\aaref@jnl{Phys.~Rev.~C}}        % Physical Review C
\def\prd{\aaref@jnl{Phys.~Rev.~D}}        % Physical Review D
\def\pre{\aaref@jnl{Phys.~Rev.~E}}        % Physical Review E
\def\prl{\aaref@jnl{Phys.~Rev.~Lett.}}    % Physical Review Letters
\def\pasp{\aaref@jnl{Publ. Astron. Soc. Pac.}}               % Publications of the ASP
\def\pasj{\aaref@jnl{PASJ}}               % Publications of the ASJ
\def\qjras{\aaref@jnl{QJRAS}}             % Quarterly Journal of the RAS
\def\skytel{\aaref@jnl{S\&T}}             % Sky and Telescope
\def\solphys{\aaref@jnl{Sol.~Phys.}}      % Solar Physics
\def\sovast{\aaref@jnl{Soviet~Ast.}}      % Soviet Astronomy
\def\ssr{\aaref@jnl{Space~Sci.~Rev.}}     % Space Science Reviews
\def\zap{\aaref@jnl{ZAp}}                 % Zeitschrift fuer Astrophysik
\def\nat{\aaref@jnl{Nature}}              % Nature
\def\iaucirc{\aaref@jnl{IAU~Circ.}}       % IAU Cirulars
\def\aplett{\aaref@jnl{Astrophys.~Lett.}} % Astrophysics Letters
\def\apspr{\aaref@jnl{Astrophys.~Space~Phys.~Res.}}
\def\bain{\aaref@jnl{Bull.~Astron.~Inst.~Netherlands}} 
\def\fcp{\aaref@jnl{Fund.~Cosmic~Phys.}}  % Fundamental Cosmic Physics
\def\gca{\aaref@jnl{Geochim.~Cosmochim.~Acta}}   % Geochimica Cosmochimica Acta
\def\grl{\aaref@jnl{Geophys.~Res.~Lett.}} % Geophysics Research Letters
\def\jcp{\aaref@jnl{J.~Chem.~Phys.}}      % Journal of Chemical Physics
\def\jgr{\aaref@jnl{J.~Geophys.~Res.}}    % Journal of Geophysics Research
\def\jqsrt{\aaref@jnl{J.~Quant.~Spec.~Radiat.~Transf.}}
\def\memsai{\aaref@jnl{Mem.~Soc.~Astron.~Italiana}}
\def\nphysa{\aaref@jnl{Nucl.~Phys.~A}}   % Nuclear Physics A
\def\physrep{\aaref@jnl{Phys.~Rep.}}   % Physics Reports
\def\physscr{\aaref@jnl{Phys.~Scr}}   % Physica Scripta
\def\planss{\aaref@jnl{Planet.~Space~Sci.}}   % Planetary Space Science
\def\procspie{\aaref@jnl{Proc.~SPIE}}   % Proceedings of the SPIE